\documentclass[lettersize,journal]{IEEEtran}
\usepackage{amsmath,amsfonts}
\usepackage{algorithmic}
\usepackage{algorithm}
\usepackage{array}
\usepackage[caption=false,font=footnotesize,labelfont=rm,textfont=rm]{subfig}
\usepackage{textcomp}
\usepackage{stfloats}
\usepackage{url}
\usepackage{verbatim}
\usepackage{graphicx}
\usepackage{cite}
\usepackage{stfloats}
\usepackage{makecell}
\usepackage{multirow}
\usepackage{amsthm,amsmath,amssymb}
\usepackage{mathrsfs}
\usepackage{epsfig}
\usepackage{epstopdf}
\usepackage{diagbox}
\usepackage{soul}
\usepackage{ulem}
\usepackage{color, xcolor}
\begin{document}

\title{The Coupling Effect of Sensing Targets on the Environment for 3GPP ISAC Channels: Observation, Modeling, and Validation}

\author{Yameng Liu, Jianhua Zhang, Yuxiang Zhang, Hongbo Xing, \\Yifeng Xiong, Zhiqiang Yuan, and Guangyi Liu
\thanks{Yameng Liu, Jianhua Zhang, Yuxiang Zhang, and Hongbo Xing are with the State Key Laboratory of Networking and Switching Technology, Beijing University of Posts and Telecommunications, Beijing 100876, China (email: liuym@bupt.edu.cn; jhzhang@bupt.edu.cn; zhangyx@bupt.edu.cn; hbxing@bupt.edu.cn).}
\thanks{Yifeng Xiong is with the School of Information and Communication Engineering, Beijing University of Posts and Telecommunications, Beijing 100876, China (email: yifengxiong@bupt.edu.cn)}
\thanks{Zhiqiang Yuan is with the National Mobile Communications Research Laboratory, Southeast University, Nanjing 210096, China (email: zqyuan@seu.edu.cn)} 
\thanks{Guangyi Liu is with China Mobile Research Institution, Beijing 100053, China (email: liuguangyi@chinamobile.com).}}

\markboth{Journal of \LaTeX\ Class Files,~Vol.~14, No.~8, August~2021}%
{Shell \MakeLowercase{\textit{et al.}}: A Sample Article Using IEEEtran.cls for IEEE Journals}


\maketitle

\begin{abstract}
Integrated Sensing And Communication (ISAC) has been identified as a key 6G application by ITU and 3GPP, with standardization efforts already underway. Sensing tasks, such as target localization, demand more precise characterization of the sensing target (ST) in ISAC channel modeling. The ST couples complexly with environmental scatterers, potentially blocking some multipaths and generating new ones, resulting in power variations compared to the original channel. To accurately model this effect, this paper proposes a coupled ISAC channel model based on measurements and validates it through similarity analysis between simulated and measured channels. In this work, we first conduct ISAC channel measurements in an indoor factory scenario at 105 GHz, where the multipath power variations caused by the ST's interaction with the environment are clearly observed. Then, we propose an ISAC channel modeling framework that incorporates two novel parameters: the Blockage-Region Coupling Factor (BR-CF) and the Forward-Scattering (FS)-CF, which characterize the spatial region and intensity of the coupling effect, respectively. 
Finally, the proposed model is validated through similarity comparison with measured data, demonstrating higher accuracy for both LoS and NLoS scenarios compared to the non-coupled model. This realistic ISAC channel model provides an effective framework for capturing the ST-environment coupling effect, supporting the design and evaluation of ISAC technologies.

\end{abstract}


\section{Introduction}\label{section1}
\IEEEPARstart{N}{owadays}, Integrated Sensing And Communication (ISAC) has been recognized as a promising technology to achieve ubiquitous sensing and digital twin for the sixth generation (6G) systems \cite{liu2020vision,liu2023shared}. The standardization efforts of ISAC has been already underway by International Telecommunication Union (ITU) and 3rd Generation Partnership Project (3GPP) \cite{3gppRan102,zhang2024latest}. Compared to conventional systems with separate devices, ISAC technology integrates the two functionalities into one system, enabling the communication Base Stations (BSs) or terminals to sense the surrounding environment \cite{kumari2017ieee}. By sharing a majority of software, hardware, and information resources, ISAC systems bring tremendous potentials of improving spectrum utilization and reducing costs \cite{liu2023joint,liu2022survey,zhang2021overview}.  

In ISAC systems, sensing tasks, such as target positioning and tracking, focus on excluding environmental clutter from the received signals and accurately extracting the effective information of the Sensing Target (ST) (e.g., delay and angle.) \cite{wang2020small,xu2023spatial}. Therefore, a realistic ISAC channel model that characterizes the impact of the ST within the propagation environment is essential, as it underpins both performance evaluation and algorithm design for ISAC systems \cite{zhang2014three,yuan2022spatial}.
In \cite{xiong2023fundamental,xiong2024torch}, all the parameters of the sensing channel are assumed to be influenced by the ST, with additive white Gaussian noise considered, and the ISAC system performance derived accordingly. Similar channel assumptions are employed in \cite{yu2022location} and \cite{chen2021code}, where beamforming and system design are explored, respectively.
However, the multipath components related to a specific ST actually constitute only a small fraction of the ISAC channel, which has been practically validated in \cite{3gppBupt118}. Treating the remaining ST-unrelated components as additive noise is an oversimplification. In \cite{gao2022toward}, a deep learning-based positioning method is proposed, assuming an idealized, denoised sensing channel that focuses solely on ST-related components, although the specifics of the model and denoising method are not provided. To accurately model the ISAC channels, the ST-related and unrelated components are defined as the target and background channels, respectively. These definitions have been agreed in the 3GPP meeting in February 2024 \cite{3gppRan116,zhang2024latest}.

\begin{figure*}[!h]
\centering
\includegraphics[width=6.8in]{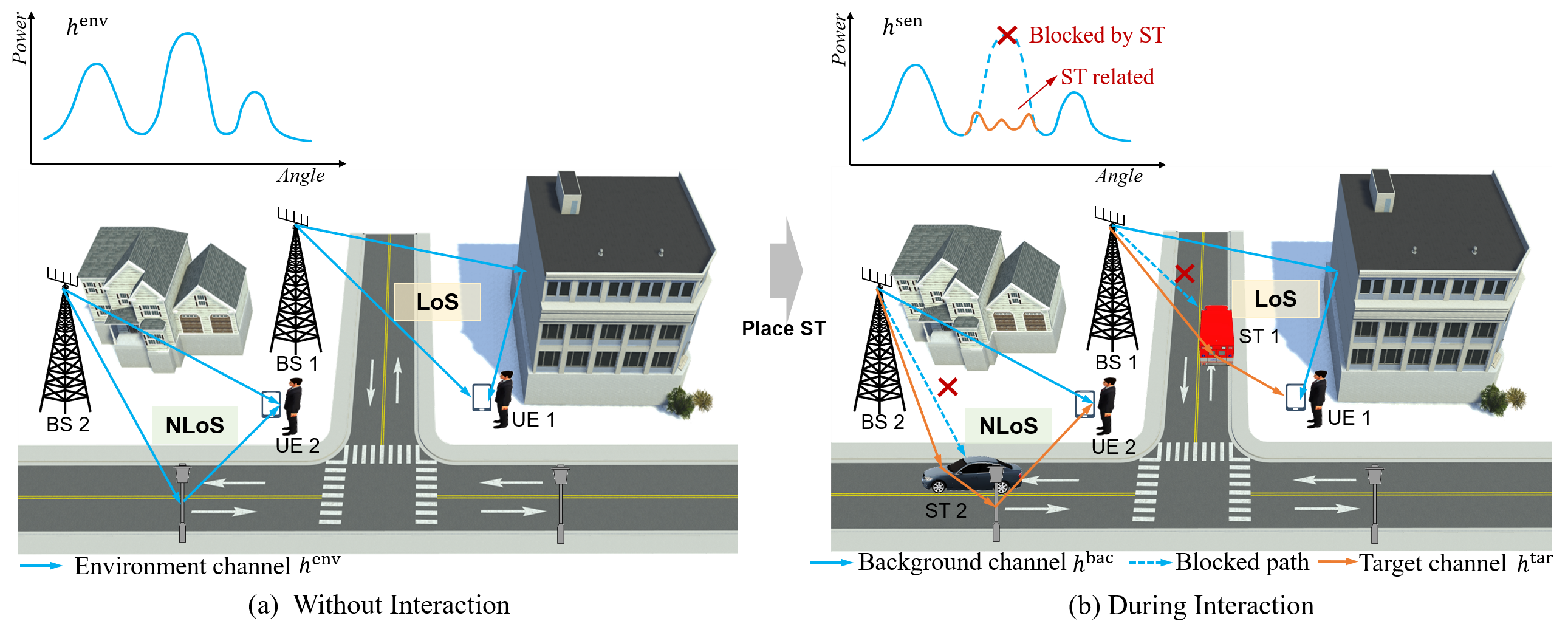}
\vspace{-0.3cm}
\caption{The illustration of ISAC system, where (a) represents the environment channel and its PAS before the ST is involved, and (b) shows the ISAC channel and its PAS with ST interaction. The orange lines denote the paths contributed by STs (i.e., vehicles), while the blue lines denote the paths contributed by environmental scatterers.}
\label{fig_sys}
\end{figure*}

The next facing challenge is how to model and combine
the target and background channels. In some existing ISAC channel studies \cite{liu2024extend,yang2024integrated,lou2023unified,luo2024channel}, the background channel is modeled as statistical clusters and directly added to the target channel components. However, this non-coupled model neglects the effect of STs on the environment, which can significantly affect modeling accuracy and result in incorrect estimation of the ST. For example, Fig. \ref{fig_sys}(a) illustrates an environment channel and its Power-Angle-Spectrum (PAS) when the ST does not interact with the propagation. When ST enters the detection range, as shown in Fig. \ref{fig_sys}(b), some original paths in the environment may be blocked, while new paths are generated through the ST. As a result, the path power under the same angle changes based on the ST’s position. Measurements in \cite{wang2022empirical} show that the closer the ST is to the transmitter (Tx) / receiver (Rx), the greater the power attenuation in the environment, indicating a stronger coupling effect. In \cite{chen2024empirical}, the large-scale coupling effect between the target and background channels is investigated through measurements, revealing that the path loss variation factor of the original channel follows a Gaussian distribution. 
Despite these findings, this issue has not been systematically considered from the perspective of channel modeling.

To address the above gaps, this paper conducts ISAC channel measurements in an Indoor Factory (InF) scenario at 105 GHz. A coupled ISAC channel modeling framework is proposed, and the similarity between the generated channel by the proposed model and the measured channel is adopted for its validation. Our major contributions and novelties are summarized as:

\begin{itemize}
\item{An ISAC channel measurement campaign with horn antenna rotation is performed in an InF scenario at 105 GHz. By analyzing the Power-Angle-Delay Profiles (PADPs) with and without the ST positioned in the environment, the coupling effect of the ISAC channel is clearly observed. Specifically, the placement of the ST induces birth-death dynamics in the multipaths, leading to power variations compared to the original channel.}

\item{A novel ISAC channel model is proposed that effectively captures the observed coupling effect of STs on the environment, enabling the combination of the target and background channels. This model introduces two parameters: the Blockage-Region Coupling Factor (BR-CF) and the Forward-Scattering (FS) -CF, which describe the spatial region and intensity of the coupling effect, respectively. Furthermore, the implementation framework of the proposed model is outlined.}

\item{Based on the proposed model, the ISAC channel is generated and validated through a similarity comparison with measured data. The BR-CF and FS-CF values extracted from Line-of-Sight (LoS) measurements align with Fresnel diffraction theory, while those under Non-LoS (NLoS) conditions follow a normal distribution. The proposed coupled model achieved approximately 30\% and 5\% higher similarity under LoS and NLoS conditions, compared to conventional non-coupled model.}
\end{itemize}

The remainder of this paper is outlined as follows. In Section \ref{section2}, descriptions of the measurements are presented and the coupling effect is observed. In Section \ref{section3}, a novel ISAC channel model is proposed, and the BR-CF and FS-CF are introduced. Then, parameterized analysis and validation of the proposed ISAC channel model are accomplished in Section \ref{section4}. Finally, Section \ref{section5} concludes the paper.

\section{Channel Measurements and Observations of the Coupling Effect}\label{section2}
\subsection{Measurement Description}

\begin{figure}[t]
\centering
\subfloat[]{\includegraphics[width=3.4in]{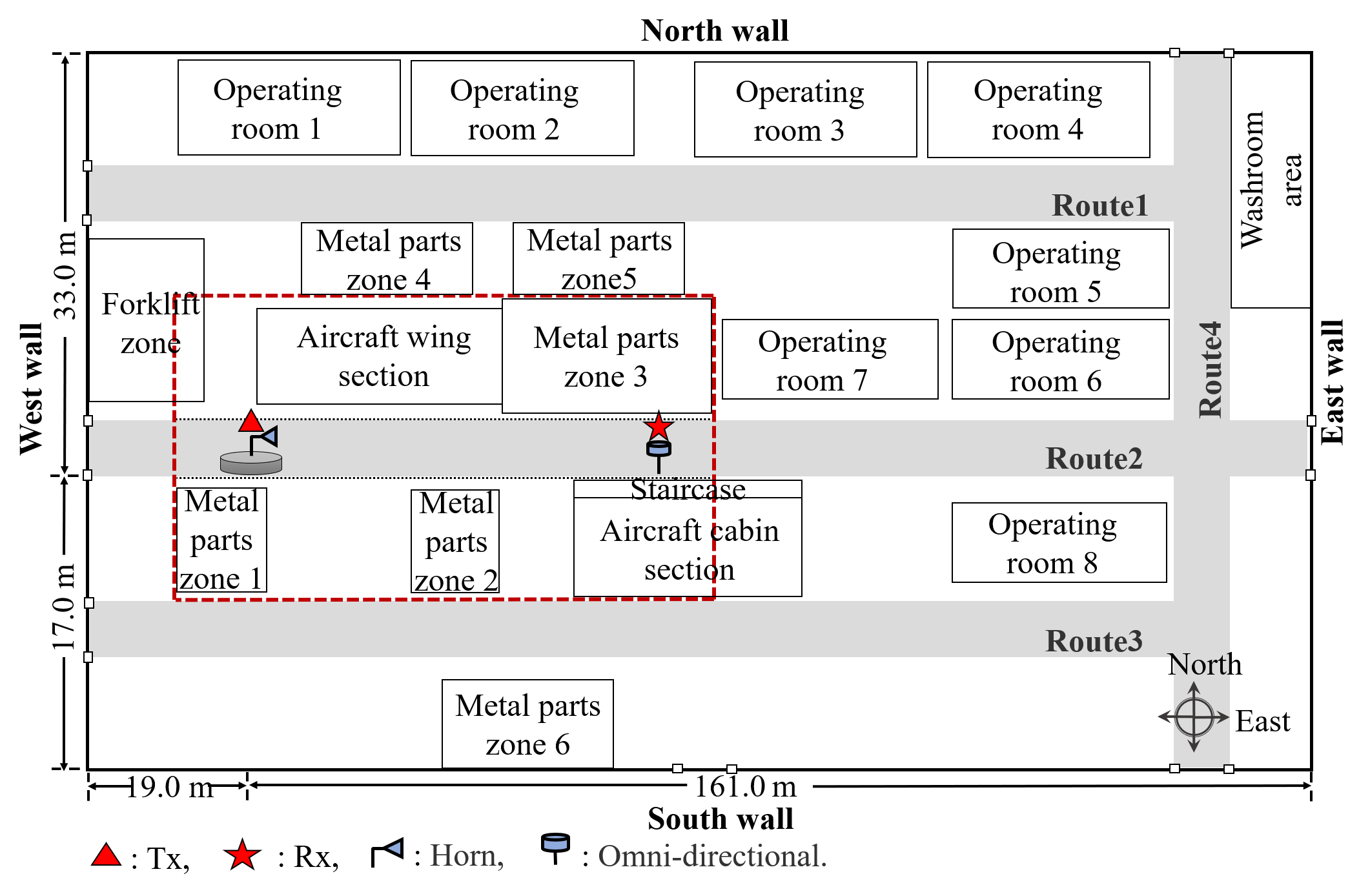}\label{}}\\
\vspace{-0.2cm}
\subfloat[]{\includegraphics[width=3.2in]{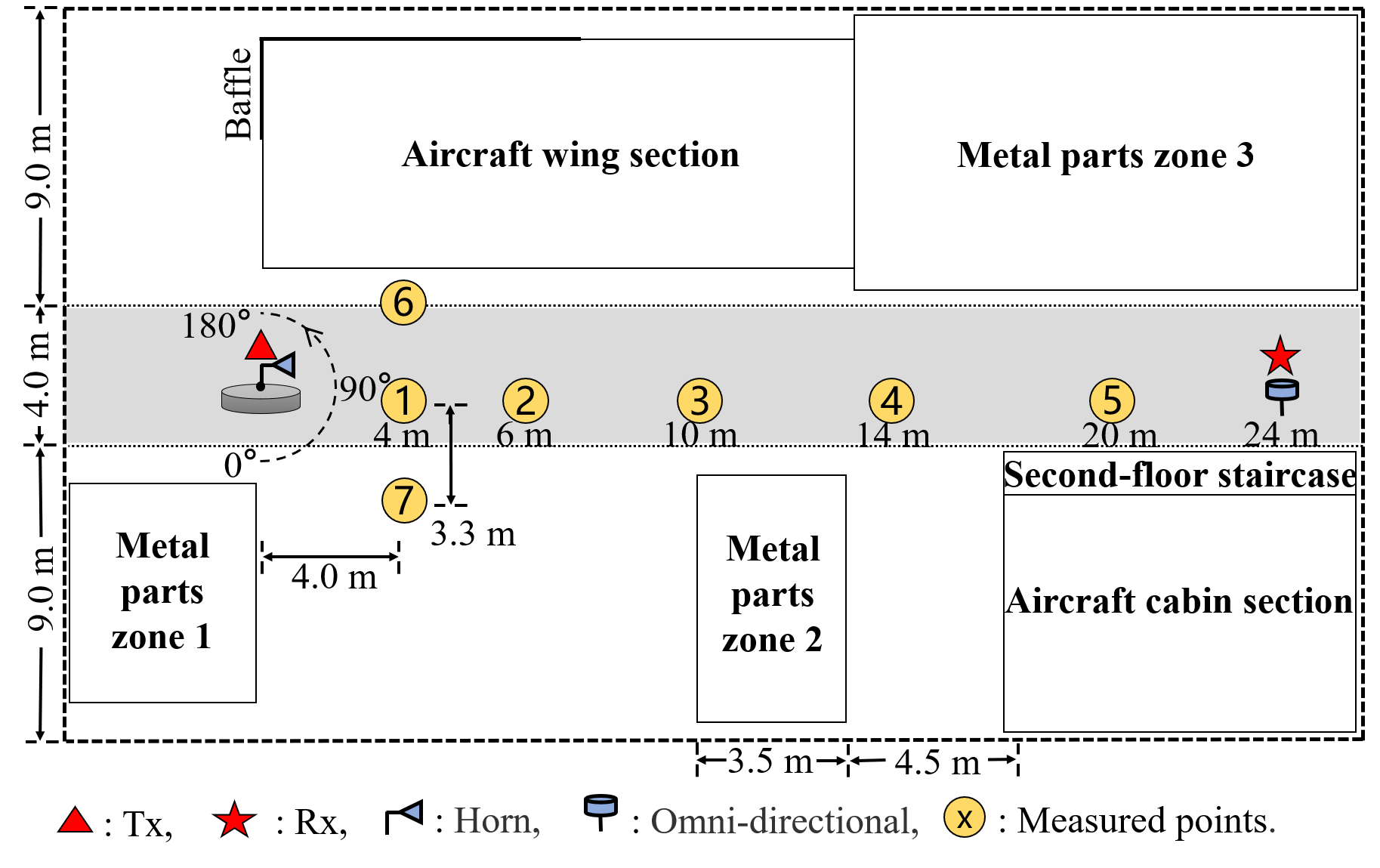}\label{}}\\
\caption{The illustration of (a) the measurement InF layout, and (b) the measurement area indicated by the red dashed borders in (a), where the yellow dots and numbers represent the Points \#1-\#7 of ST. }
\label{fig_sce1}
\end{figure}

\begin{figure}[h]
\centering
\subfloat[]{\includegraphics[width=2.8in]{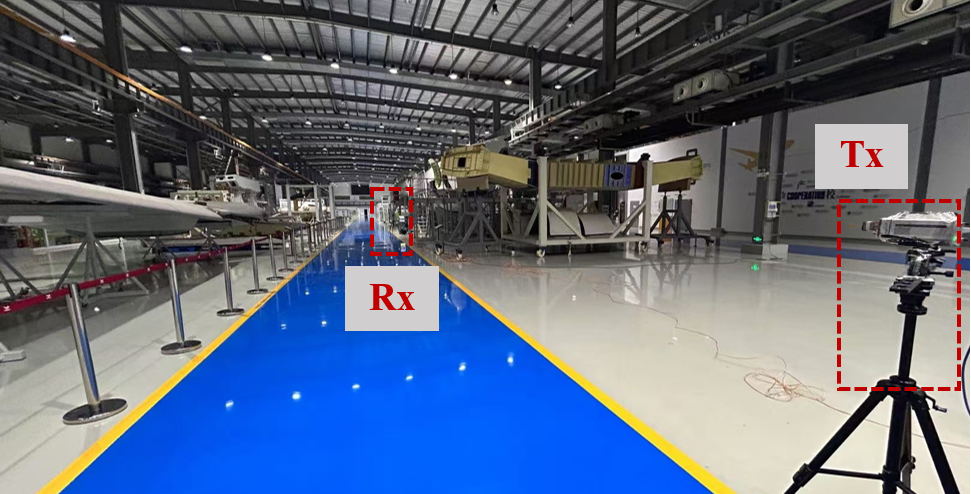}\label{}}\\
\vspace{-0.2cm}
\subfloat[]{\includegraphics[width=2.8in]{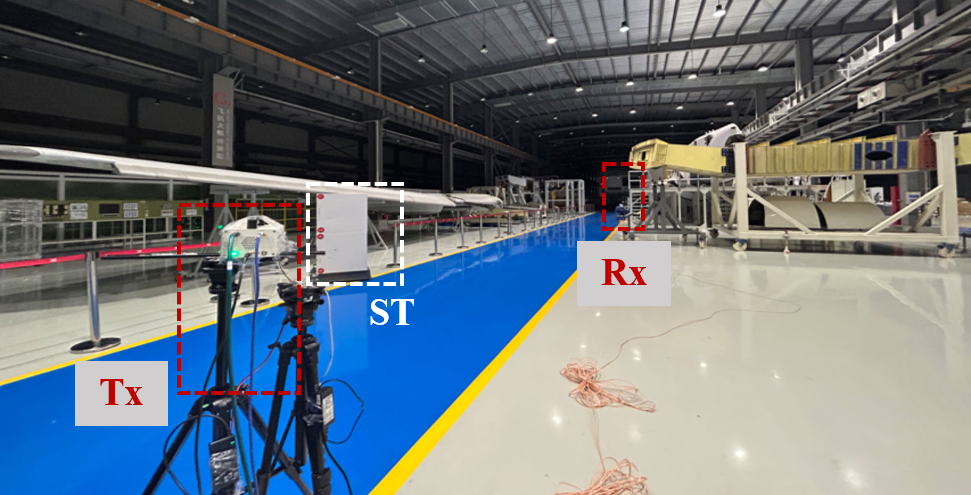}\label{}}
\caption{The measurement photographs taken for (a) the environment channel and (b) the ISAC channel at measured Point \#6.}
\label{fig_sce2}
\end{figure}

The channel measurement campaign is conducted on an internal road in the InF scenario at 105 GHz, using a loaded Automated Guided Vehicle (AGV) as the ST, surrounded by airplane wings and metal components. The measurement layout and realistic surroundings are illustrated in Fig. \ref{fig_sce1} and Fig. \ref{fig_sce2}.
During the measurements, a horn antenna is configured at the Tx side (indicated by the red triangle in Fig. \ref{fig_sce1}) and horizontally rotated with a step of 5° from south to north, covering a range of 180°. 
An omni-directional antenna is used at the Rx side (marked by the red pentagram in Fig. \ref{fig_sce1}), located 24 m away from the Tx, to receive multipaths scattered by the environment.
Initially, the environment channel without the detection ST is measured, referred to as Point \#0. Subsequently, the AGV, loaded with two metal boxes and standing at a total height of 1.6 m, is placed at 7 positions (referred to as measured Points \#1-\#7), as indicated by yellow dots in Fig. \ref{fig_sce1}(b). Points \#1-\#5 are located at the LoS condition between the Tx and Rx, while Points \#6 and \#7 are under NLoS condition. Point \#6 is surrounded by abundant scatterers, whereas Point \#7 is relatively open.


A wideband channel sounder operating at THz frequencies is exploited to extract the ISAC channel characteristics. At the Tx side, a vector signal generator (R\&S SMW 200A) generates a Pseudo Noise (PN) sequence with a code rate of 500 Msym/s and a length of 2047. The baseband PN sequence is modulated into an Intermediate Frequency (IF) signal at 12 GHz using Binary Phase Shift Keying (BPSK) modulation. The signal has a zero-to-zero bandwidth of 1 GHz. The IF signal is then modulated upon Local Oscillator (LO) signal generated by a signal generator (R\&S SMB 100A) using a frequency multiplier, producing a probing signal centered at 105 GHz. To ensure a satisfactory receiving Signal-to-Noise Ratio (SNR), the signal is amplified by a power amplifier with 43 dB gain before being transmitted via a high-gain horn antenna.
At the Rx side, a spectrum analyzer (R\&S FSW 43) is utilized to demodulate the THz signal received by the antenna. A low-noise amplifier with a typical gain of 20 dB is applied. At each measured point, 4094 I/Q samples are collected at a sample rate of 1 GHz, corresponding to a delay resolution of 1 ns. Absolute propagation delay is obtained by synchronizing the spectrum analyzer and the signal generator via direct cable connection. Finally, the Channel Impulse Responses (CIRs) of the measurement scenarios are derived through data processing on the laptop. Detailed parameters of the channel measurement configuration are summarized in Table. \ref{table_1}.

\begin{table}[t]
\caption{The Channel Measurement Configuration\label{table_1}}
\centering
\begin{tabular}{cc}
\hline
Parameter & Values\\
\hline
ST size & 0.7$\times$0.6$\times$1.6 m$^3$ \\
Points \#1-\#5 condition & LoS \\
Points \#6 and \#7 condition & NLoS \\
Central frequency & 105 GHz\\
IF / OF frequency & 12 / 31 GHz\\
Symbol rate & 500 Msym/s\\
Bandwidth & 1 GHz\\
PN sequence & 2047\\
Sampling rate & 1 GHz\\
Tx / Rx antenna type & Horn / Omni\\
Horn antenna azimuth / vertical HPBW  & 8.5$^\circ$ / 9.9$^\circ$\\
Omni antenna azimuth / vertical HPBW  & 360$^\circ$ / 30$^\circ$\\
Horn antenna gain & 25 dBi\\
Omni-directional antenna gain & 3 dBi\\
Antenna height & 1.4 m\\
\hline
\end{tabular}
\end{table}

\subsection{Channel Observations} 

\begin{figure*}[!h]
\centering
\includegraphics[width=6in]{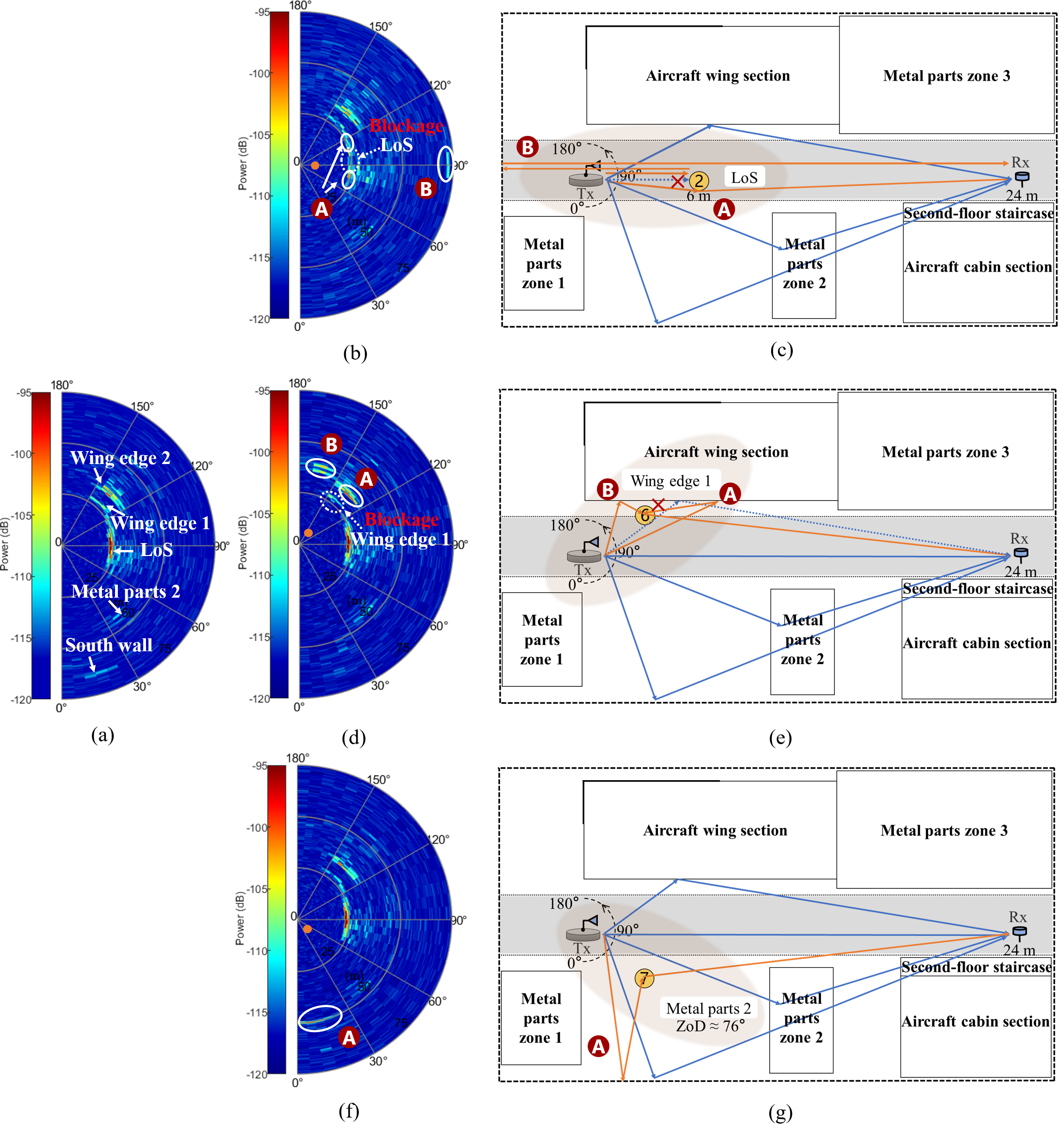}
\vspace{-0.3cm}
\caption{The measured results. (a) PADP for environment channel (before ST interaction). (b) PADP and (c) propagation route illustration when ST is positioned at the LoS point (Point \#2). (d) PADP and (e) propagation route illustration when ST is positioned at the NLoS point in a scatterer-rich environment (Point \#6). (f) PADP and (g) propagation route illustration when ST is positioned at the NLoS point in an open environment (Point \#7). In (c), (e), and (g), the orange lines denote paths contributed by ST and the blue lines denote paths contributed only by environmental scatterers.}
\label{fig_padp}
\end{figure*}

To remove the system response from the equipment and cables, we perform a back-to-back calibration on the field measurement data. The CIR at each ST point for all rotation angles is measured as $h(\theta,\tau)$, where $\tau$ and $\theta$ denote the propagation delay and rotation angle, respectively. The PADP of the channel can be calculated using the following formula
\begin{equation}
\label{eqn_1}
{\text{PADP}(\theta,\tau)}=\left| h(\theta,\tau) \right|^2.
\end{equation}

In Fig. \ref{fig_padp}, the semicircular images (i.e., (a)(b)(d)(f)) illustrate the measured PADPs, where the center represents the Tx position, and the radius corresponds to the propagation distance of the paths. Considering the detection distance, only paths within 75 m are depicted here, which occupies more than 95\% of the energy. The angles of the semicircles represent the path Angles of Departure (AoDs), approximated by the rotation angles of the horn antenna. The depth of color represents the magnitude of the received power (dB). Simplified scenario diagrams (i.e., Fig. \ref{fig_padp}(c)(e)(g)) illustrate the potential propagation routes corresponding to the PADPs (b)(d)(f), with background channel paths in blue and target channel paths in orange. Blocked paths are indicated by dashed lines and white boxes.

This paper uses results from four measured points as examples: Point \#0 (environment channel without ST interaction), Point \#2 (Tx-Rx LoS condition), and Points \#6 and \#7 (NLoS condition). Fig. \ref{fig_padp}(a) shows the environmental PADP. 
The dominant LoS paths are observed at a AoD of 90° and a distance of 24 m, reflecting the Rx location. This paths account for approximately 88.93\% of the total power. By matching the actual layout, other single-bounce propagation paths from scatterers (e.g., wing edges, metal parts, walls) are also identified in Fig. \ref{fig_padp}(a). 

\begin{figure}[h]
\centering
\vspace{-0.3cm}
{\includegraphics[width=3.0in]{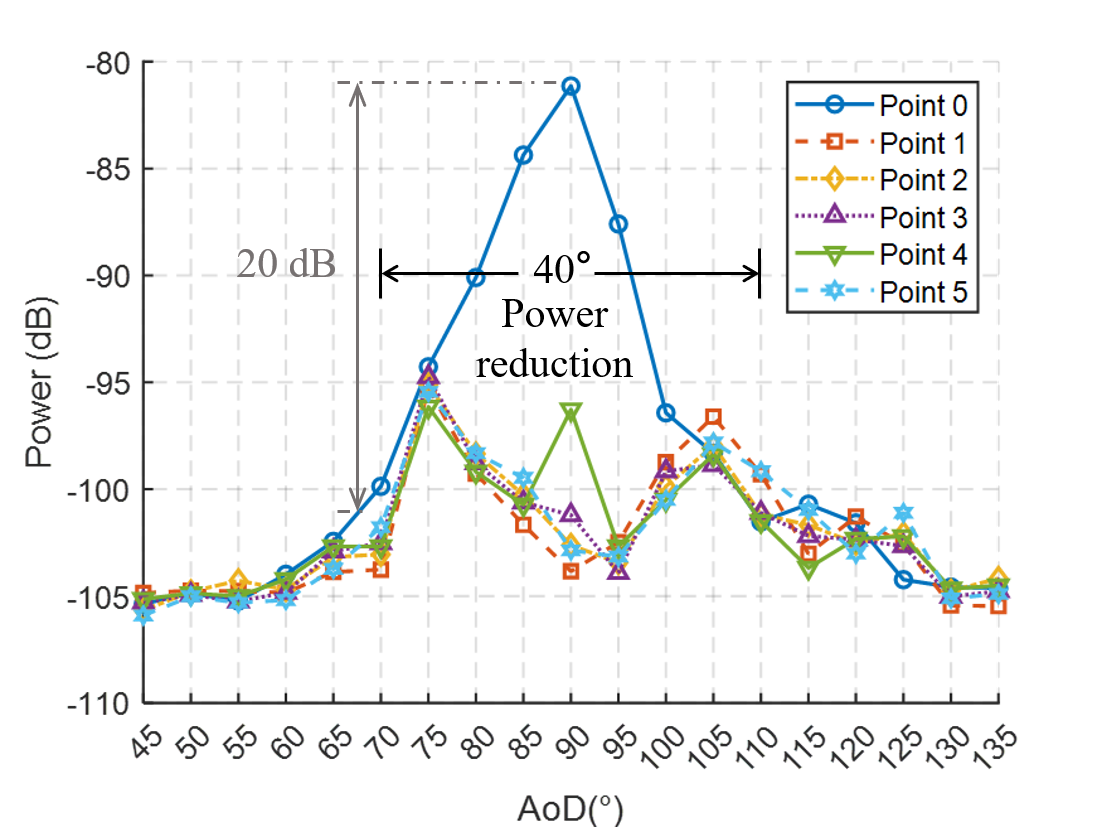}
}
\vspace{-0.2cm}
\caption{The PAS around LoS paths of the environment channel (Point \#0) and measured Points \#1-\#5. The propagation delay ranges from 75 to 89 ns (22.5 to 26.7 m).}
\vspace{-0.2cm}
\label{fig_result1}
\end{figure}

The PADP of Point \#2 is presented in Fig. \ref{fig_padp}(b), where the ST position is indicated by an orange dot. The angle from Tx to the ST center is 90°, with the ST's angular size approximately 6.7°. Fig. \ref{fig_padp}(c) illustrates the potential propagation routes corresponding to (a) and (b). 
In Fig. \ref{fig_padp}(b), the LoS path (Tx-Rx) is blocked, indicated by a dotted circle. Simultaneously, Tx-ST-Rx paths (classified as direct paths), arising from scattering or diffraction at the ST's horizontal edges, are labeled as red letter \textit{A}. The PAS around these paths is shown in Fig. \ref{fig_result1}, where the introduction of the ST (Points \#1-\#5) causes a power attenuation of approximately 20 dB in directions ranging from 70° to 110° (a 40° span). Notably, the power increase at Point \#4 at 90° results from the ST being positioned midway between the Tx and Rx. This position causes minimal obstruction and allows direct propagation or diffraction over the ST’s upper edge.
Moreover, non-coupled indirect paths, labeled \textit{B}, are illustrated in Fig. \ref{fig_padp}(b), which are formed along Tx-ST-west wall-ST-Rx. 
The propagation routes of the above newborn paths \textit{A} and \textit{B} in the target channel are highlighted with orange lines in Fig. \ref{fig_padp}(c). The target channel accounts for 37.38\% of the total channel power, with 91.23\% of its power contributed by the direct paths \textit{A}.

\begin{figure}[h]
\centering
\vspace{-0.3cm}
{\includegraphics[width=3.0in]{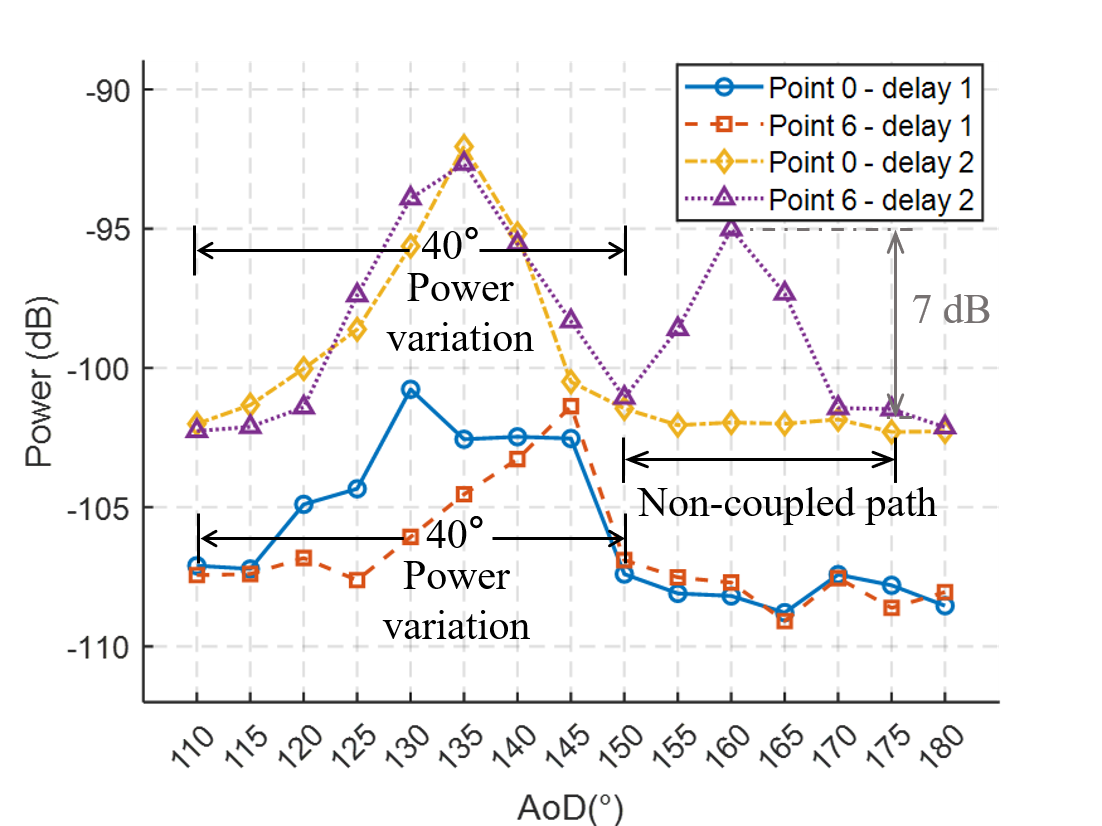}
}
\vspace{-0.2cm}
\caption{The PAS around NLoS paths of the environment channel (Point \#0) and measured Points \#6. The propagation delay 1 ranging from 88 to 94 ns (26.4 to 28.2 m) near the propagation scatterer of wing edge 1, while the delay 2 ranging from 107 to 134 ns (38.7 to 40.2 m) near wing edge 2 and 3. }
\label{fig_result2}
\end{figure}

When the ST is located at measured Point \#6, the angle from Tx to the ST center is approximately 130°, with the the ST's angular size around 7.8°. The PADP and the corresponding propagation routes are illustrated in Fig. \ref{fig_padp}(d) and (e). Comparing (d) with Fig. \ref{fig_padp}(a), path variations are observed around 130°, with some paths (e.g., those propagating via Tx-wing edge1-Rx) being blocked, as indicated by the dotted circle in (d). Meanwhile, newly generated coupled paths, marked by solid a circle and red letter \textit{A}, and non-coupled paths around 160° labeled as \textit{B} , represent target channel components. Scenario analysis reveals that paths labeled \textit{A} (around 130°) and \textit{B} (around 160°) are indirect paths propagating via Tx-ST-wing edge2-Rx and Tx-wing edge3-ST-Rx, respectively. These paths contribute 5.60\% and 1.66\% of the total sensing channel power, with their routes are highlighted by orange lines in Fig. \ref{fig_padp}(e).
The PAS around these paths is shown in Fig. \ref{fig_result2}, where power variations are observed approximately in directions ranging from 110° to 150° (a 40° span). The non-coupled generated paths exhibit peak power increases of 7 dB compared to the original paths.

\begin{figure}[h]
\centering
\vspace{-0.3cm}
{\includegraphics[width=3.0in]{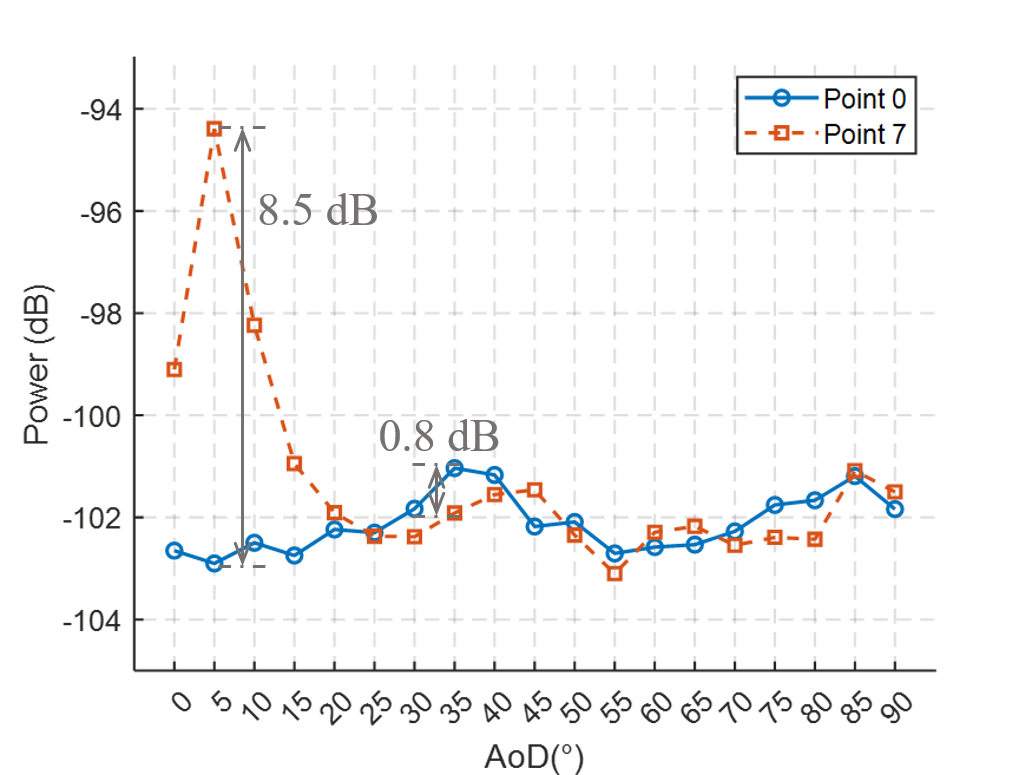}
}
\vspace{-0.2cm}
\caption{The PAS around NLoS paths of the environment channel (Point \#0) and measured Point \#7. The propagation delay ranges from 145 to 170 ns (43.5 to 51.0 m).}
\label{fig_result4}
\end{figure}

At Point \#7, the angle from the Tx to the ST center is approximately 50°, with the ST's angular
size also around 7.8°. Fig. \ref{fig_padp}(f) and (g) illustrate the PADP and potential propagation routes. Comparing (f) with Fig. \ref{fig_padp}(a), no significant coupling changes are observed for the environmental multipaths, as the environment near Point \#7 is relatively open. 
Note that the weak path Tx-metal parts 2-Rx, as shown in Fig. \ref{fig_padp}(g), originates from the upper edge of a metal component approximately 3 m in height, with a Zenith Angle of Departure (ZoD) about 76°. Therefore, the placement of the ST does not result in noticeable blockage effects, as the channel measurements in this study are conducted only at a ZoD of 90°. In Fig. \ref{fig_result4}, within the range of 20° to 90°, the maximum power variation between Point \#0 and Point \#7 is 0.8 dB, primarily influenced by noise floor fluctuations. Additionally, within the range of 0° to 15°, newly formed (non-coupled) paths show a peak power increase of 8.5 dB. These paths, labeled \textit{A}, propagate via Tx-north wall-ST-Rx, as illustrated in Fig. \ref{fig_padp}(f) and (g).

\subsection{Discussion}

\begin{figure*}[!h]
\centering
\includegraphics[width=4.95in]{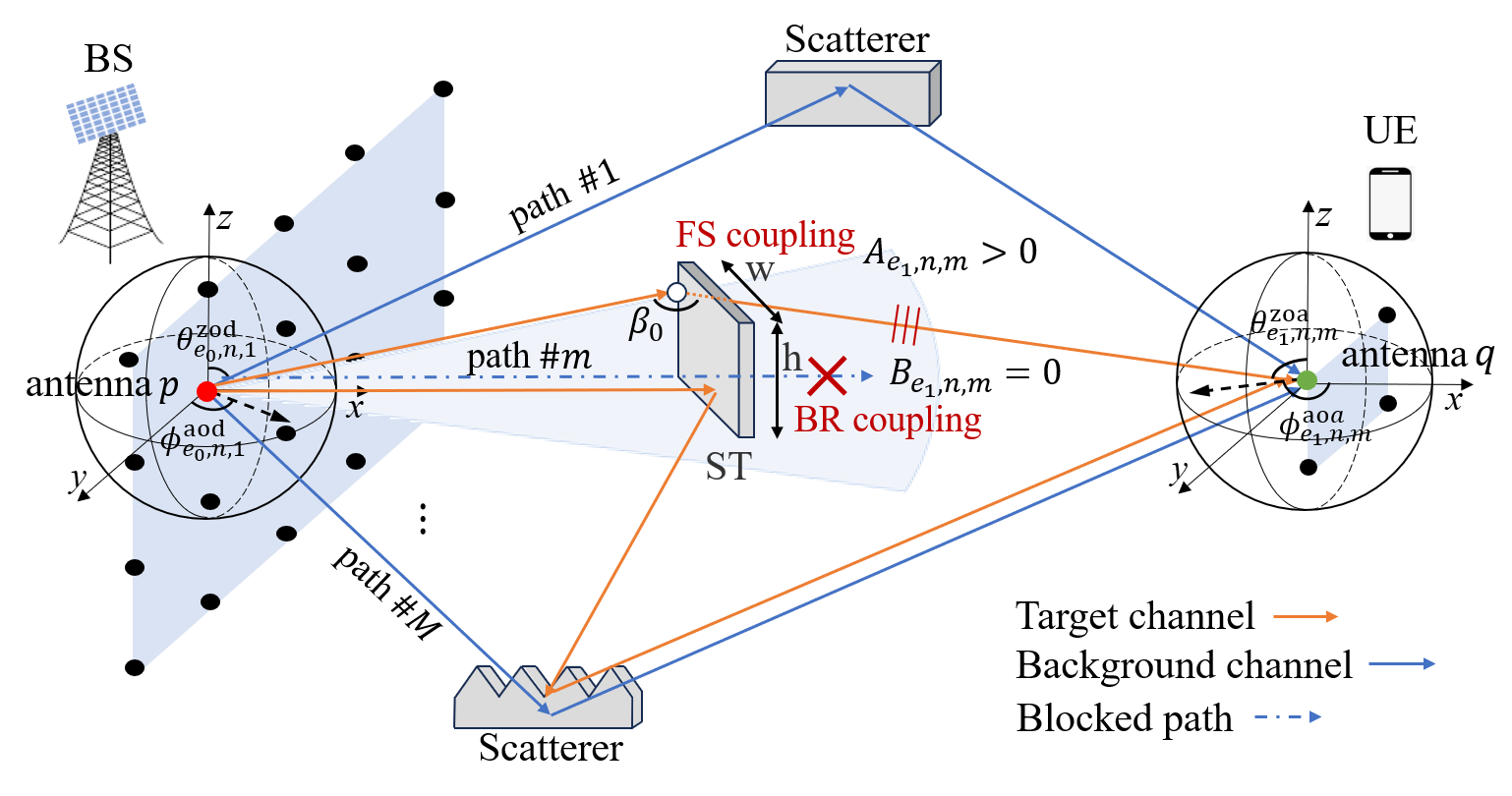}
\vspace{-0.3cm}
\caption{The illustration of ISAC channel model with coupling effect between ST and the environment. The blue lines denote the environment / background channel paths, and the orange lines denote the target channel paths. The blue dashed lines indicate blocked paths.}
\label{fig_model}
\end{figure*}

Based on the above analysis, we conclude that deploying an ST (here, an AGV loaded with metal boxes) into the environment (here, an InF scenario) induces the following interesting coupling effect in the ISAC channel:

\begin{itemize}
    \item \textit{Reduction of environment paths:} when the ST is positioned along the Tx-Rx LoS or Tx-scatterer-Rx NLoS paths, the corresponding paths in the environment channel attenuate or disappear. The affected region, centered around the ST, is larger than its angular size and influenced by the practical beamwidth. In our measurements, this region spans approximately 40° in the horizontal angular domain under both LoS and NLoS conditions.
\end{itemize}

\begin{itemize}
    \item \textit{Generation of target paths:} within the affected region, paths originating from the ST’s edges may emerge. These newborn, coupled paths in the target channel exhibit delays and angles similar to the original ones. Outside this region, non-coupled generated paths are distributed within a broader Angular Spread (AS) around the ST. Moreover, target paths can be direct or indirect. Direct paths, when present, tend to dominate the power contribution in the target channel. 
\end{itemize}

\section{ISAC Channel Model with Coupling Effect of STs on the Environment}\label{section3}

This paper considers the ISAC channel under frequency-selective fading. 
The BS and User Equipment (UE) are equipped with $[1,...p,...,P]$ and $[1,...,q,...,Q]$ antennas, respectively. In the downlink propagation scenario, the BS transmits and the UE receives the signal. The ISAC channel model between antenna $p$ and $q$ is illustrated in Fig. \ref{fig_model}, with the propagation scatterers and the ST represented as grey geometric objects. The blue and orange lines illustrate the paths in the background and target channels, respectively. When an ST is positioned along the BS-UE line, the original LoS paths, present in the absence of the ST, becomes blocked (indicated by the dotted blue lines). The spatial affected region where this blockage occurs is defined as Blockage-Region (BR). Simultaneously, radio waves may bypass the ST edges and reach the UE through scattering or diffraction, forming target channel paths. The concept of Forward-Sattering\footnote{The scattering direction of the object for electromagnetic waves aligns closely with the incident direction \cite{skolnik1970radar}, i.e., $\beta_0$ in Fig. \ref{fig_model} approaches 180°.} (FS), which is widely used in radar research, can be applied to describe these coupled paths. 
Moreover, non-coupled paths in the target channel (e.g., BS-ST-scatterer-UE) are also depicted with orange lines in Fig. \ref{fig_model}.

\subsection{Channel Modeling}

This paper extends the Geometry-Based Stochastic channel Model (GBSM) and incorporates the coupling effect in ISAC channel. GBSM has been widely adopted by ITU and 3GPP standards \cite{3gpp38901}, which is a clustered structure model that groups paths into stochastic clusters. We define the state of environment channel (before ST interaction) as $e_0$. The CIR between Tx antenna $p$ and Rx antenna $q$ at time $t$ is expressed as: 
\begin{equation}
\label{eqn_env}
{h_{q,p,{e_0}}^{\rm{env}}(t,\tau ) =  \!\sum\limits_{{n_{{e_0}}}}^{{N_{{e_0}}}} \! \sum\limits_{{m_{{e_0}}}}^{{M_{{e_0}}}} \!{h_{q,p,{e_0},n,m}^{\rm{env}}(t)}\delta(\tau - \tau_{e_0,n,m}^{q,p}),} 
\end{equation}
where $\tau$ represents the propagation delay. ${h_{q,p,{e_0},n,m}^{\rm{env}}(t)}$ denotes the complex coefficients for path $m$ in cluster $n$ at the state of $e_0$. $n_{e_0}=1,2, \ldots ,{N_{e_0}}$ and $m_{e_0}=1,2, \ldots ,{M_{e_0}}$ are the indexes of the clusters and paths in environment channel. 

When the ST interacts with the environment, the ISAC channel state is defined as $e_1$. Considering the coupling effect between ST and the environment, the ISAC channel response between $p$ and $q$ at time $t$ is given by:
\begin{equation}
\label{eqn_isac1}
{h^{\rm{sen}}_{q,p,e_1}}(t,\tau ) = {h^{\rm{tar}}_{q,p,e_1}}(t,\tau ) + {h^{\rm{bac}}_{q,p,e_1}}(t,{\tau}),
\end{equation}
\begin{equation}
\label{eqn_isac2}
h_{q,p,{e_1}}^{\rm{bac}}(t,\tau )= \mathop{\underline{{{h}}_{q,p,{e_0}}^{\rm{env,1}}(t,\tau )}}\limits_{{\rm{blocked}}} + \mathop{\underline{{{h}}_{q,p,{e_0}}^{\rm{env,2}}(t,\tau )}}\limits_{{\rm{unaffected}}},
\end{equation}
\begin{equation}
\label{eqn_isac3}
h_{q,p,{e_1}}^{\rm{tar}}(t,\tau )= \mathop{\underline{{{h}}_{q,p,{e_1}}^{\rm{tar,1}}(t,\tau )}}\limits_{{\rm{coupled}}} + \mathop{\underline{{{h}}_{q,p,{e_1}}^{\rm{tar,2}}(t,\tau )}}\limits_{{\rm{non-coupled}}},
\end{equation}
where ${h_{q,p,{e_1}}^{\rm{tar}}(t,\tau )}$ and ${h_{q,p,{e_1}}^{\rm{bac}}(t,\tau )}$ denote the CIRs between $p$ and $q$ of target and background channels at the state of $e_1$, respectively. 

In (\ref{eqn_isac2}), certain paths in the environment channel are coupled / blocked (${{h}}_{q,p,{e_0}}^{\rm{env,1}}(t,\tau )$) based on the position of ST, while the second term (${{h}}_{q,p,{e_0}}^{\rm{env,2}}(t,\tau )$) represents paths in the environment that remain unaffected. To model this effect, we firstly propose a BR Coupling Factor (BR-CF) in the background channel to characterize the ST's BR. Accordingly, (\ref{eqn_isac2}) is expressed as
\begin{equation}
\label{eqn_bac}
h_{q,p,{e_1}}^{\rm{bac}}(t,\tau )= {{\bf{B}}_{q,p,{e_1}}}(t) \cdot {\bf{h}}_{q,p,{e_0}}^{\rm{env}}(t,\tau ),
\end{equation}
where ${\bf{h}}_{{q,p,e_0}}^{\rm{env}}(t,{\tau})=[{{h}}_{{q,p,e_0,n,m}}^{\rm{env}}(t)\delta({\tau})]_{({N_{e_0} \times M_{e_0}})\times 1}$ denotes the vector form of environment CIR with dimensions $({N_{e_0} \times M_{e_0}})\times 1$. ${{\bf{B}}_{q,p,{e_1}}(t)}=[{{B}}_{{e_1},n,m}(t)]_{1\times({N_{e_0} \times M_{e_0}})}$ represents the BF-CF with dimensions $1\times({N_{e_0} \times M_{e_0}})$ at the state of $e_1$, where the element for path $m$ in cluster $n$ is a Boolean variablecan, denoted as ${{{B}}_{{e_1},n,m}(t)}=0\ \rm{or}\ 1$. These two parameters can both be defined as the dimensions of $1\times{N_{e_0}}$ when applied at the cluster level.

In (\ref{eqn_isac3}), ${{h}}_{q,p,{e_1}}^{\rm{tar,1}}(t,\tau )$ and ${{h}}_{q,p,{e_1}}^{\rm{tar,2}}(t,\tau )$ represent the channel response of newly generated paths that within and outside the BR, respectively. ${{h}}_{q,p,{e_1}}^{\rm{tar,2}}(t,\tau )$ can be calculated by a concatenation of Tx-ST and ST-Rx links \cite{zhang2024cascaded}, expressed as
\begin{equation}
\small
\label{eqn_con}
{{h}}_{q,p,{e_1}}^{\rm{tar,2}}(t,\tau ) = {{h}}_{q,l,{e_1}}^{\rm{tar,2}}(t,\tau,\Gamma_{\rm{out}} )*{h_{l,p,e_1}^{\rm{tar,2}}}(t,{\tau},\Gamma_{\rm{in}})\cdot{\sigma _l}(\Gamma_{\rm{out}},\Gamma_{\rm{in}}),
\end{equation}
where $l$ denotes the index of the STs. ${{h}}_{q,l,{e_1}}^{\rm{tar,2}}(t,\tau,\Gamma_{\rm{out}} )$ and ${h_{l,p,t_1}^{\rm{tar,2}}}(t,{\tau},\Gamma_{\rm{in}})$ represent the non-coupled channel responses for the ST-Rx and Tx-ST links, respectively. $*$ indicates the convolution operation. ${\sigma}_l(\Gamma_{\rm{out}},\Gamma_{\rm{in}})$ represents the composite RCS values of the ST, characterizing the fading of the path power induced by the ST \cite{skolnik1970radar}. The RCS values are determined by the outgoing ($\Gamma_{\rm{out}}=\{\theta _{{e_1},n,m}^{{\rm{out}}},\phi _{{e_1},n,m}^{{\rm{out}}}\}$) and incident angles ($\Gamma_{\rm{in}}=\{\theta _{{e_1},n,m}^{{\rm{in}}},\phi _{{e_1},n,m}^{{\rm{in}}}\}$) associated with the ST \cite{liu2024extend}. 

Note that the coupled ${{h}}_{q,p,{e_1}}^{\rm{tar,1}}(t,\tau )$ is related to the original environmental paths with similar delays and angles. To accurately model the power effect of the ST on the environment, this paper proposes an FS-CF, providing a more generalized representation of ${{h}}_{q,p,{e_1}}^{\rm{tar,1}}(t,\tau )$ as:
\begin{equation}
\label{eqn_tar1}
h_{q,p,{e_1}}^{\rm{tar,1}}(t,\tau ) = {{\bf{\tilde B}}_{q,p,{e_1}}}(t) \cdot {{\bf{A}}_{q,p,{e_1}}}(t) \cdot {\bf{h}}_{q,p,{e_0}}^{\rm{env}}(t,\tau ),
\end{equation}
where ${{\bf{\tilde B}}_{q,p,{e_1}}}(t)$ represent the inverse transformation of ${{\bf{B}}_{q,p,{e_1}}}(t)$. ${{\bf{A}}_{q,p,{e_1}}(t)}=[{{A}}_{{e_1},n,m}(t)]_{1\times({N_{e_0} \times M_{e_0}})}$ represents the FS-CF with dimensions $1\times({N_{e_0} \times M_{e_0}})$ at the state of $e_1$, where the element for path $m$ in cluster $n$ is ${{A}}_{{e_1},n,m}(t)\geq0$. This parameter can be defined as the dimensions of $1\times{N_{e_0}}$ when applied at the cluster level. In open environments, the FS-CF can also be interpreted as an RCS value at a specific angle (i.e., when $\beta_0$ approaches 180°).

\begin{figure*}[b]
\begin{small}
\begin{subequations}\label{eqn_model}
\begin{align}
\hline \notag\\
\label{eqn_modele0}
h_{q,p,{e_0}}^{\rm{env}}(t,\tau ) = &\sum\limits_{{n_{{e_0}}}}^{{N_{{e_0}}}}  \sum\limits_{{m_{{e_0}}}}^{{M_{{e_0}}}} {a_{{e_0},n,m}^{q,p}{{ {{F_{rx,q} (\theta _{{e_0},n,m}^{{\rm{zoa}}},\phi _{{e_0},n,m}^{{\rm{aoa}}})} } }}  } {{F_{tx,p} (\theta _{{e_0},n,m}^{{\rm{zod}}},\phi _{{e_0},n,m}^{{\rm{aod}}})}} \exp \left( {\frac{{j2\pi (\hat r_{{e_0},rx,n,m}^T \cdot {{\bar d}_q})}}{{{\lambda _0}}}} \right)  \notag\\ 
& \cdot \exp \left( {\frac{{j2\pi (\hat r_{{e_0},tx,n,m}^T \cdot {{\bar d}_p})}}{{{\lambda _0}}}} \right)\exp \left( {j2\pi f_{d,{e_0},n,m}^{q,p}t} \right) {\exp (j\Phi _{{e_0},n,m})} \delta (\tau  - \tau _{{e_0},n,m}^{q,p}),\\
\label{eqn_modele1}
h_{q,p,{e_1}}^{\rm{sen}}(t,\tau ) = &\sum\limits_{{n_{{e_1}}}}^{N_{{e_1}}^{\rm{t2}}}  \!\sum\limits_{{m_{{e_1}}}}^{M_{{e_1}}^{\rm{t2}}} \!{a_{{e_1},n,m}^{q,l}a_{{e_1},n,m}^{l,p}{{{
{F_{rx,q} (\theta _{{e_1},n,m}^{{\rm{zoa}}},\phi _{{e_1},n,m}^{{\rm{aoa}}})}
} }}  } {
{\sigma _l(\Gamma^{\rm{out}}_{e_1,n,m},\Gamma^{\rm{in}}_{e_1,n,m})}}   {
{F_{tx,p} (\theta _{{e_1},n,m}^{{\rm{zod}}},\phi _{{e_1},n,m}^{{\rm{aod}}})}}  \exp \! \left( \!{\frac{{j2\pi (\hat r_{{e_1},rx,n,m}^T \cdot {{\bar d}_q})}}{{{\lambda _0}}}} \! \right) \notag\\
& \cdot \exp \left( {\frac{{j2\pi (\hat r_{{e_1},tx,n,m}^T \cdot {{\bar d}_p})}}{{{\lambda _0}}}} \right) \exp \left( {j2\pi (f_{d,{e_1},n,m}^{q,l} + f_{d,{e_1},n,m}^{l,p})t} \right) {
{\exp (j\Phi _{{e_1},n,m})}} \delta (\tau  - \tau _{{e_1},n,m}^{q,l} - \tau _{{e_1},n,m}^{l,p}) \notag\\
& + \sum\limits_{{n_{{e_0}}}}^{{N_{{e_0}}}}  \sum\limits_{{m_{{e_0}}}}^{{M_{{e_0}}}}  ({{\tilde B}_{{e_1},n,m}(t)} \cdot {A_{{e_1},n,m}(t)} + {B_{{e_1},n,m}(t)}) \cdot h_{q,p,{e_0},n,m}^{\rm{env}}(t)\delta (\tau  - \tau _{{e_0},n,m}^{q,p}).
\end{align}
\end{subequations}
\end{small}
\end{figure*}

Based on the aforementioned discussion, the impulse responses of the environment channel $h_{q,p,{e_0}}^{\rm{env}}(t,\tau )$ in (\ref{eqn_env}) and the ISAC channel $h_{q,p,{e_1}}^{\rm{sen}}(t,\tau )$ in (\ref{eqn_isac1}) based on GBSM are expressed as (\ref{eqn_model}). The specific notation is detailed as follows:
\begin{itemize}
\item{$\lambda_0$ is the wavelength of the carrier frequency.}
\item ${a_{{e_0},n,m}^{q,p}}$ denote the amplitude for path $m$ in cluster $n$ between antenna $q$ and $p$ at the state of $e_0$. Similarly, ${a_{{e_1},n,m}^{q,l}}$ and ${a_{{e_1},n,m}^{l,p}}$ denote the amplitudes for path $m$ in cluster $n$ of ST $l$ to antenna $q$ and antenna $p$ to ST $l$ links at the state of $e_1$, respectively. 
\item ${\phi _{{e_0},n,m}^{{\rm{aod}}}}$, ${\phi _{{e_0},n,m}^{{\rm{aoa}}}}$, ${\phi _{{e_1},n,m}^{{\rm{aod}}}}$, and ${\phi _{{e_1},n,m}^{{\rm{aoa}}}}$ represent the AoD and Azimuth angle of Arrival (AoA) for path $m$ in cluster $n$ at the state of $e_0$ and $e_1$, respectively.
\item ${\theta _{{e_0},n,m}^{{\rm{zod}}}}$, ${\theta _{{e_0},n,m}^{{\rm{zoa}}}}$, ${\theta _{{e_1},n,m}^{{\rm{zod}}}}$, and ${\theta _{{e_1},n,m}^{{\rm{zoa}}}}$ represent the ZoD and zenith angle of arrival (ZoA) for path $m$ in cluster $n$ at the state of $e_0$ and $e_1$, respectively.
\item{$F_{rx,q}$ and $F_{tx,p}$ are the radiation patterns of the Rx antenna $q$ and Tx antenna $p$.}
\item{$\Phi_{{e_0},{n},{m}}$ and $\Phi_{{e_1},{n},{m}}$ respectively denote the random initial phases for path $m$ in cluster $n$.}
\item{$\kappa _{{e_0},{n},{m}}$ and $\kappa _{{e_1},{n},{m}}$ represent the cross polarization power ratios (XPRs) for path $m$ in cluster $n$ at the state of $e_0$ and $e_1$, respectively.}
\item{$\widehat r_{{e_0},rx,{n},{m}}$, $\widehat r_{{e_0},tx,{n},{m}}$, $\widehat r_{{e_1},rx,{n},{m}}$, and $\widehat r_{{e_1},tx,{n},{m}}$ are the spherical unit vectors of path $m$ in cluster $n$ at Rx and Tx sides for the state of $e_0$ and $e_1$, respectively.}
\item{${\overline d }_{q}$ and ${\overline d }_{p}$ are the location vectors of antennas $q$ and $p$.}
\item{$f_{d,{e_0},{n},{m}}$ and $f_{d,{e_1},{n},{m}}$ denote the Doppler shifts for path $m$ in cluster $n$ at the state of $e_0$ and $e_1$, respectively.}
\item {$\tau _{{e_0},n,m}^{q,p}$ represents the delay for path $m$ in cluster $n$ at the state of $e_0$. $\tau _{{e_1},n,m}^{q,l}$ and $\tau _{{e_1},n,m}^{l,p}$ represent the delays for path $m$ in cluster $n$ of ST $l$ to antenna $q$ and antenna $p$ to ST $l$ links at the state of $e_1$, respectively.}
\item {$n_{e_1}^{\rm{t2}}=1,2,...,N_{e_1}^{\rm{t2}}$ is the cluster indexes of non-coupled target channel $h_{q.p.e_1}^{\rm{tar,2}}$. $m_{e_1}^{\rm{t2}}=1,2,...,M_{e_1}^{\rm{t2}}$ denotes the corresponding path indexes.}
\end{itemize}
For clarity and to focus on the coupling effect, (\ref{eqn_model}) is simplified to consider single polarization. Nonetheless, this model remains applicable to dual-polarization scenario.

\begin{figure}[h]
\centering
\vspace{-0cm}
{\includegraphics[width=2.8in]{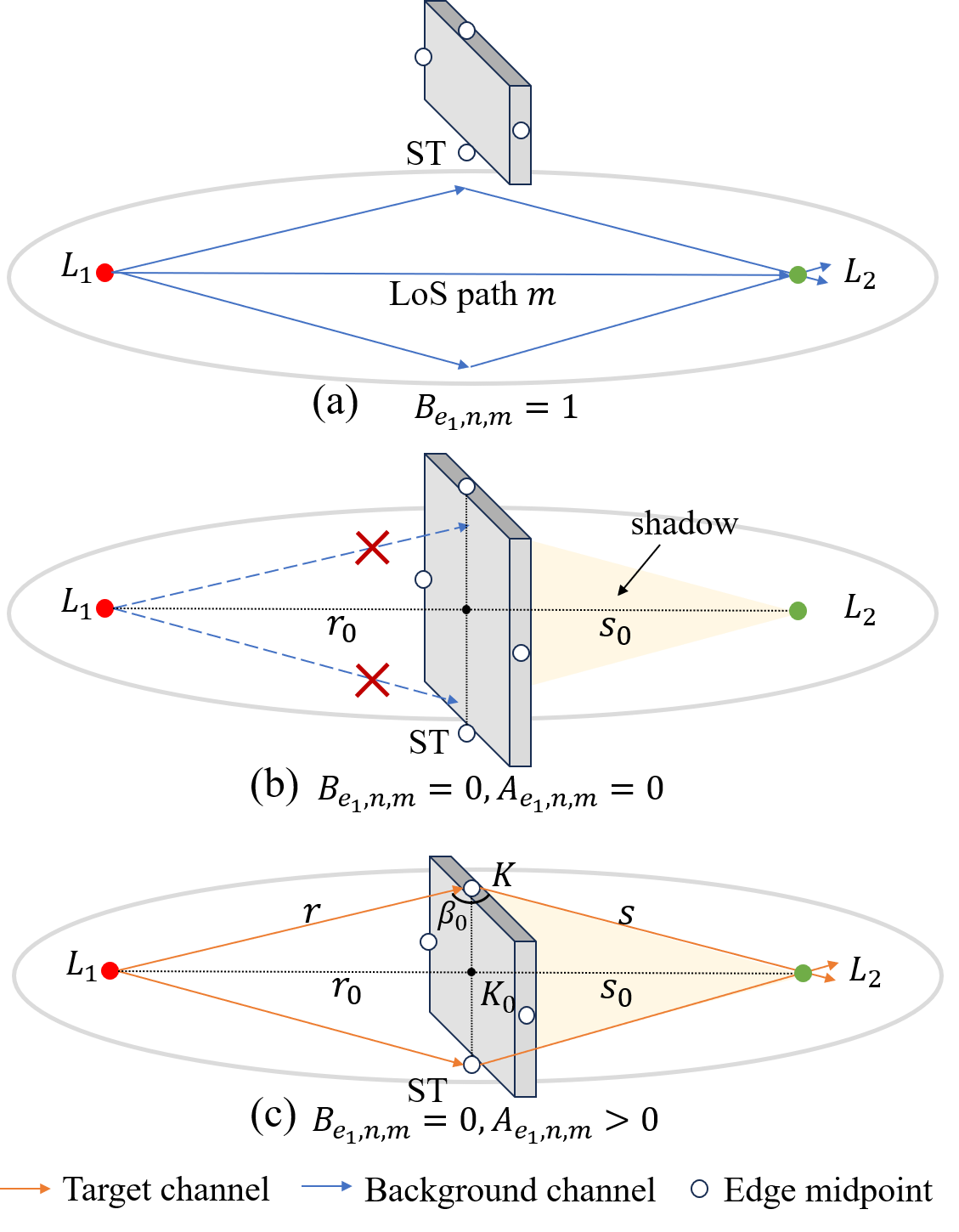}
}
\vspace{-0cm}
\caption{The detailed illustration of BR-CF and FS-CF.}
\label{fig_ked}
\end{figure}

In (\ref{eqn_model}), BR-CF $B_{e_1,n,m}(t)$ equals 0 when the environmental path lies within the ST's BR and 1 otherwise. The BR size, influenced by the ST and antenna beamwidth, is larger than the angular size of the ST (as observed in Section \ref{section2}) and encompasses all delays behind the ST in the time domain. The FS-CF, $A_{e_1,n,m}(t)$, has a valid value only when $B_{e_1,n,m}(t)=0$. Detailed FS-CF illustrations based on BR-CF are provided in Fig. \ref{fig_ked}.

\textit{1) No Interaction} (Fig. \ref{fig_ked}(a)): Path $m$ in cluster $n$ between transmission nodes $L_1$ and $L_2$ lies outside the BR ($B_{e_1,n,m}=1$), and the ISAC channel can be considered as a direct superposition of the target and background channels (e.g., Point \#7 in Section \ref{section2}). Here, the target channel does not include any components coupled to the environmental paths, so $h_{q,p,{e_1}}^{\rm{tar,1}}(t,\tau ) = 0$. 

\textit{2) Within BR but no FS} (Fig. \ref{fig_ked}(b)): Path $m$ lies within the BR ($B_{e_1,n,m} = 0$). Consequently, the corresponding environmental path is deleted. However, due to the large size of the ST, no propagation path associated with the original $L_1-L_2$ path is generated, resulting in $A_{e_1,n,m} = 0$.

\textit{3) Within BR and with FS} (Fig. \ref{fig_ked}(c)): $B_{e_1,n,m} = 0$. Electromagnetic waves bypass the edges of the ST, undergoing scattering or diffraction. This generates new coupled paths, leading to $A_{e_1,n,m} > 0$. 

As for $A_{e_1,n,m}$ in case 3), the primary electromagnetic propagation involves diffraction and scatterering occurring in open environments, along with potential interactions with surrounding scatterers. 
To analyze the theoretical values of FS-CF, this paper considers an ST with widths $w_1$ and $w_2$, and heights $h_1$ and $h_2$ in an open environment. Taking point $K$ on the ST integration surface as an example, as shown in Fig. \ref{fig_ked}(c), the projection of $K$ onto the line from $L_1$ to $L_2$ is denoted as $K_0$. The distances from $L_1$ and $L_2$ to $K_0$ are $r_0$ and $s_0$, while the distances from $L_1$ and $L_2$ to $K$ are $r$ and $s$, respectively. Assuming $\beta_0 \approx 180^\circ$ and the ST integration surface is perpendicular to the $L_1L_2$ line, the electromagnetic field received at $L_2$ can be derived using the Fresnel-Kirchhoff diffraction formula \cite{born2013principles}:
\begin{equation}
\label{eqn_fk}
U=-\frac{E_0}{\lambda} \frac{i}{r_0 s_0} \iint_{\mathcal{S}} e^{i k(r+s)} d S,
\end{equation}
where $k = 2 \pi/ \lambda$. $E_0$ denotes the electric field strength at a unit
distance in free space. $S$ represents the integration region.  

Given the excessive complexity of (\ref{eqn_fk}), the Knife-Edge Diffraction (KED) model—a commonly used blockage model that treats the obstruction as a perfectly absorbing thin plate—can be adopted to simplify the calculation. Communication channel standards such as TR 38.901 \cite{3gpp38901} apply a simplified 4KED model, considering only the impact of four edge center points of the obstruction on propagation. 
In this simplified model, the diffraction attenuation, which can serve as FS-CF values, for the path $m$ in cluster $n$ is expressed as:
\begin{equation}
\label{eqn_ked}
{{A}}_{{e_1},n,m}[\rm{dB}]=-20 \log _{10}\left(1-\left(F_{h_1}+F_{h_2}\right)\left(F_{w_1}+F_{w_2}\right)\right),
\end{equation}
where $F_{h_1}$, $F_{h_2}$, $F_{w_1}$, and $F_{w_2}$ account for KED at the four edges corresponding to the heights and widths of the ST. For a single edge, the result can be further approximated using an arc-tangent function, avoiding complex numerical integration: 
\begin{equation}
\label{eqn_fi}
F_i=\frac{U}{U^0}=\frac{1}{\pi}{\tan^{-1}\left( \pm \frac{\pi}{2} \sqrt{\frac{\pi}{\lambda}\left(o_i \cdot (r+s)-r_0-s_0\right)}\right)},
\end{equation}
where $U^0=\frac{E_0 e^{ik(r_0+s_0)}}{(r_0+s_0)}$ is the amplitude in free space. $i=h_1|h_2|w_1|w_2$. $o_i$ denote the sign function. For paths not intersecting the ST in a side view, $o_{h_1|h_2} = -1$ and $o_{w_1|w_2} = 1$; for those not intersecting in a top view, $o_{w_1|w_2} = -1$ and $o_{h_1|h_2} = 1$ \cite{3gpp38901}.
Considering the antenna gain, the FS-CF in dB can be calculated by the 4KED-G model as (\ref{eqn_kedg}) \cite{kizhakkekundil2021four}. In this formula, $G_{r_{h_1|h_2|w_1|w_2}}$ and $G_{s_{h_1|h_2|w_1|w_2}}$ denote the linear gains of the Tx and Rx antennas for the four edges, respectively.

\begin{figure}[h]
\centering
\vspace{-0.3cm}
{\includegraphics[width=3.4in]{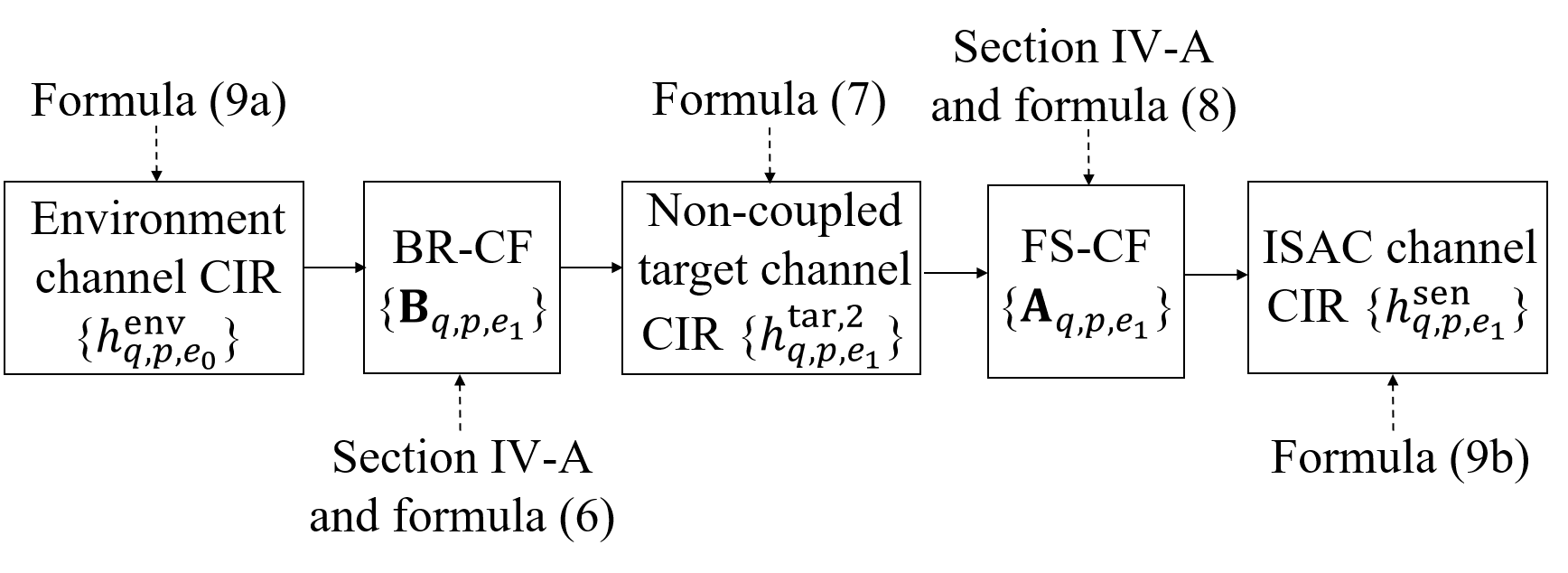}}
\vspace{-0.1cm}
\caption{The implementation framework of the proposed ISAC channel modeling with the coupling effect.}
\label{fig_frame}
\end{figure} 

\begin{figure*}
\begin{equation}
\begin{aligned}
\label{eqn_kedg}
{{A}}_{{e_1},n,m}^{\rm{G}}[{\rm{dB}}]=& -20 \log _{10}\left[1-\left[1-\left[\left(\frac{1}{2}-F_{h 1}\right) \sqrt{G_{r_{h 1}}} \sqrt{G_{s_{h 1}}}+\left(\frac{1}{2}-F_{h 2}\right) \sqrt{G_{r_{h 2}}} \sqrt{G_{s_{h 2}}}\right]\right]\right. \\ 
& \left.\times\left[1-\left[\left(\frac{1}{2}-F_{w 1}\right) \sqrt{G_{r_{w 1}}} \sqrt{G_{s_{w 1}}}+\left(\frac{1}{2}-F_{w 2}\right) \sqrt{G_{r_{w 2}}} \sqrt{G_{s_{w 2}}}\right]\right]\right].\\ \\
\hline 
\end{aligned}
\end{equation}
\end{figure*}

\subsection{Channel Implementation Framework}

Here, we present how to generate a realistic ISAC channel based on the proposed coupled channel model. Fig. \ref{fig_frame} illustrates the implementation framework. Taking the channel between antennas $p$ and $q$ as an example, the environment CIR (${h}_{q,p,e_0}^{\rm{env}}(t,\tau)$) without ST interaction is first generated using (\ref{eqn_env}) and (\ref{eqn_modele0}) in Section \ref{section3}-A. The modeling methodology and parameters in communication standards \cite{3gpp38901} can be effectively reused. Then, the BR-CF value ($\textbf{B}_{q,p,e_1}(t)$) can be generated by considering the size and different positions of the ST in the environment. BR-CF can either be applied to individual paths within a cluster or simplified to affect the entire cluster. In the third step of the channel realization, the non-coupled paths of the target channel ($h_{q,p,e_1}^{\rm{tar,2}}(t,\tau)$) outside BR is generated based on (\ref{eqn_con}) and the first component in (\ref{eqn_model}b). In the forth step, FS-CF ($\textbf{A}_{q,p,e_1}(t)$) are applied to $\textbf{h}_{q,p,e_0}^{\rm{env}}(t,\tau)$ to describe the coupled power variances within BR compared to the original paths, which form part of the target channel. The specific model parametrization will be introduced in the following Section \ref{section4}-A. Finally, the ISAC channel response ($h_{q,p,e_1}^{\rm{sen}}(t,\tau)$) is generated according to (\ref{eqn_model}b), completing the modeling process.

\section{Model Parametrization and Validation}\label{section4}

In this section, we firstly extract the BR-CF and FS-CF values from measurements for LoS and NLoS points. These values are fitted based on the theoretical analysis in \ref{section3}-A and applied to the proposed model. Next, we introduce a Similarity Index (SI) to compare the channel generated by the conventional target and background non-coupled and the proposed coupled model with the measured ISAC channel, thereby validating the accuracy of the proposed model.

\subsection{The Parameterization of BF-CF and FS-CF}

At the measured Point \#7 in Section \ref{section2}, no significant coupling effect between the ST and the environment is observed, as most original paths lie outside the BR, i.e., $\textbf{B}_{q,p,e_1} = \textbf{I}$. This sub-section focuses on the environment channel and measured Points \#1-\#6 to extract BF-CF and FS-CF values. It is worth noting that, we only consider the angle dimension within a predefined delay range. This simplification aligns with the 3GPP standards, which defines intra-cluster paths as having the same delay.

\textit{1) ST coupling with environmental LoS paths}

We analyze the environment channel (Point \#0) and ISAC channel when the ST is positioned along the LoS paths between Tx and Rx (Points \#1-\#5). The environmental scatterers around the 90° LoS paths are sparse, and the involved electromagnetic propagation relatively straightforward. As shown in Fig. \ref{fig_result1} of Section \ref{section2}-B, the BR is defined as 70° to 110°. Within this range, the BR-CF values are set to 0, while values less than 70° or greater than 110° are set to 1.

It is important to note that the horn antenna used in the measurements has an azimuth beamwidth of 8.5°. Although the typical antenna gain values have been excluded from the data analysis, the results presented in Fig. \ref{fig_result1} inherently include the impact of the antenna radiation pattern due to its non-omnidirectional characteristics.
To exclude the influence of the antenna pattern and derive the pure FS-CF values at each AoD, we calculate the power ratio between Points \#1-\#5 and the environment channel in the linear domain. The FS-CF calculation can be expressed as
\begin{equation}
\begin{aligned}
\label{eqn_fsm}
A_{{e_1}}^{\rm{mea}}[\rm{dB}](\phi) = & 10{\rm{log}}_{10}\frac{{P}_{e_1}(\phi)}{P_{e_0}(\phi)}\\
= & 10{\rm{log}}_{10}\frac{{\sum\limits_\tau  {{{\left| {a_{{e_1},n,m}^{q,p}(\tau ,\phi)} \right|}^2}} }}{{\sum\limits_\tau  {{{\left| {a_{{e_0},n,m}^{q,p}(\tau ,\phi)} \right|}^2}} }},
\end{aligned}
\end{equation}
where $A_{{e_1}}^{\rm{mea}}[\rm{dB}](\phi )$ is the measured FS-CF value at the angle of $\phi $ in the dB domain. ${P}_{e_1}(\phi)$ and $P_{e_0}(\phi )$ denote the multipath total power at $\phi $ within a preset delay range. ${a_{{e_1},n,m}^{q,p}(\tau ,\phi )}$ and ${a_{{e_0},n,m}^{q,p}(\tau ,\phi )}$ represent the amplitudes of paths at delay $\tau$ and angle $\phi $ for the states $e_1$ and $e_0$, respectively.

\begin{figure}[h]
\centering
{\includegraphics[width=3.2in]{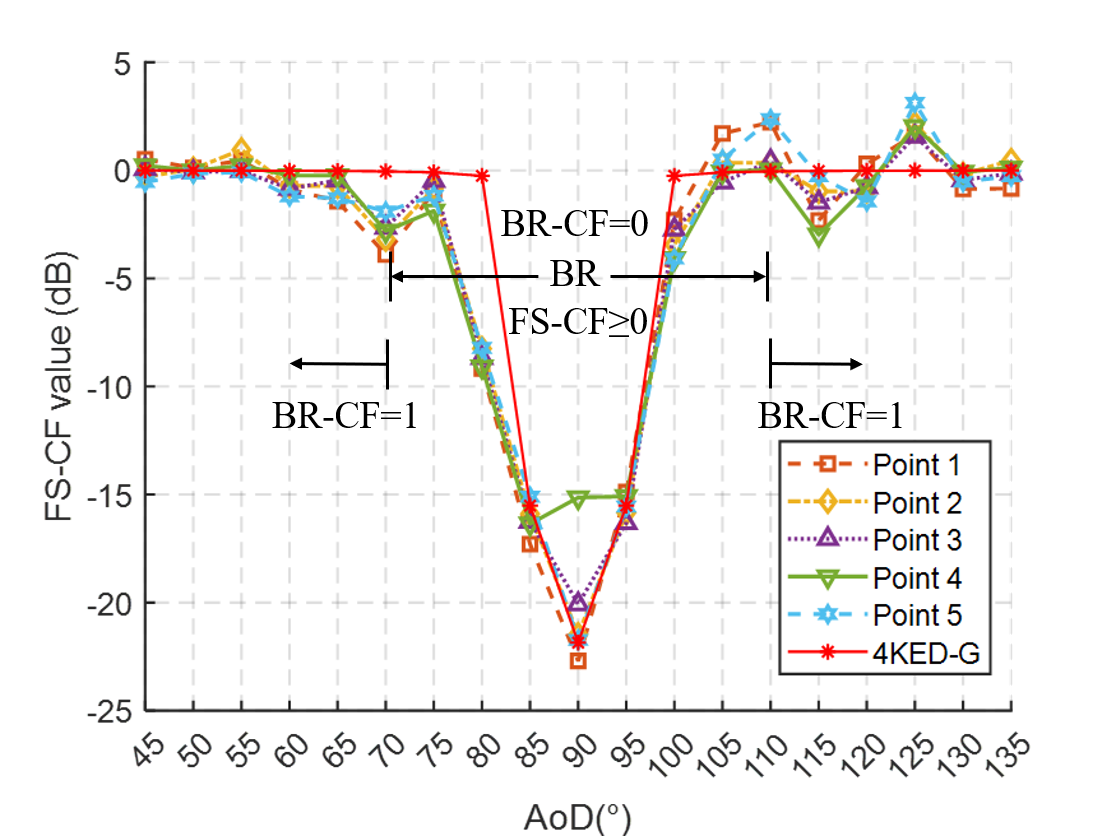}}
\caption{The results of BF-CF and FS-CF values extracted from measurements in the angular domain, when ST is interacted with environmental LoS paths, i.e., Points \#1-\#5. The red curve denote the result derived by the theoretical 4KED-G model in (\ref{eqn_kedg}).}
\label{fig_result1b}
\end{figure}

As shown in Fig. \ref{fig_result1b}, the calculated FS-CF values indicate significant fading in the LoS paths within the BR. With the horn antenna gradually rotating towards 90°, the FS-CF value decreases from approximately 0 to about -20 dB, with noticeable jitter near the BR edges. The asymmetry in the FS-CF distribution is attributed to the hollow metal staircase near the AoD angle of less than 90°, as shown in Fig. \ref{fig_sce1}(b), where complex reflections and scattering occur due to the non-ideal open environment.

Based on (\ref{eqn_fk})-(\ref{eqn_kedg}) in Section \ref{section3}-A, we model the FS-CF at each AoD by equivalently translating the ST instead of rotating the horn antenna, ensuring that the reference amplitude $U^0$ in (\ref{eqn_fi}) remains constant. In the simulation, the heights of the Tx and Rx, their separation, and the size of the ST are configured according to Table \ref{table_1}. Taking Point \#1, where the ST is positioned 4 m from the Tx along the Tx-Rx propagation, as an example, the FS-CF is calculated using (\ref{eqn_kedg}). The measured data aligns well with the theoretical 4KED-G model (indicated by the red curve in Fig. \ref{fig_result1b}), with only minor deviations observed at angles smaller than 90°.

\textit{2) ST coupling with environmental NLoS paths}

When the ST interacts with environmental NLoS paths (Point \#6), the environmental scatterers around the AoD of 130° are abundant. As discussed in Section \ref{section2}-B, significant multipath variations are primarily distributed around this angle and wthin two delay ranges. As shown in Fig. \ref{fig_result2}, the BR is defined between 120° to 150°. Within this range, the BR-CF values are set to 0, while the values less than 120° or greater than 150° are set to 1. Once the BR-CF is determined, the background channel can be generated according to (\ref{eqn_isac2}). 

\begin{figure}[h]
\centering
\vspace{-0.2cm}
\subfloat[]{\includegraphics[width=3.2in]{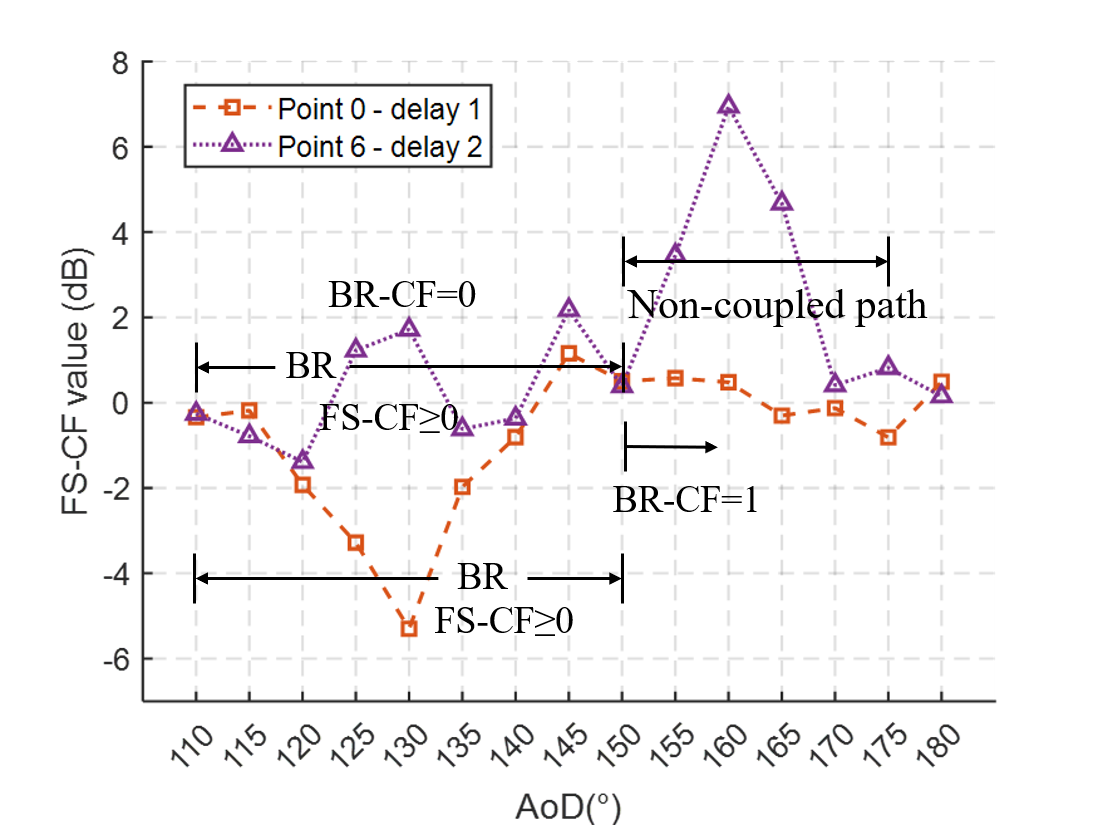}}\\
\vspace{-0.2cm}
\subfloat[]{\includegraphics[width=3.2in]{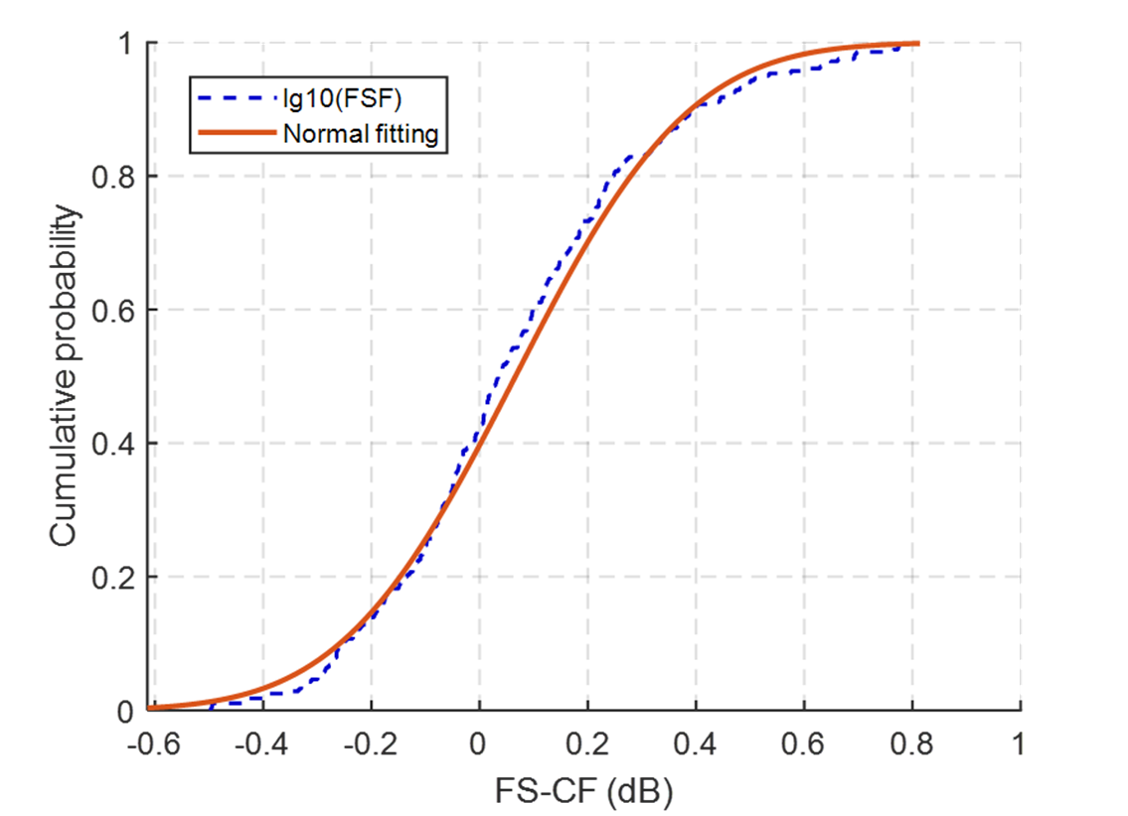}}
\vspace{-0cm}
\caption{The (a) BR-CF and FS-CF values and (b) CDF of FS-CF values extracted from measurements in the angular domain, when ST is interacted with environmental NLoS paths, i.e., Point \#6.}
\label{fig_result2bc}
\end{figure}

Fig. \ref{fig_result2bc}(a) illustrates the FS-CF values extracted from measurements at Point \#6, calculated using (\ref{eqn_fsm}). In delay 1, non-coupled multipath components contributed by the ST and the environmental scatterers exhibit between 150° and 170°, outside the BR. Within BR for both delay 1 and 2, the complexity of the environment leads to various propagation, causing the FS-CF values to deviate from the theoretical Fresnel model in this case. 
To further model the FS-CF values, we conduct a statistical analysis on 20 measured snapshots across the BR, covering two delay ranges, resulting in 360 samples (9 angles $\times$ 2 ranges $\times$ 20 snapshots). The Cumulative Distribution Function (CDF) of the results is presented in Fig. \ref{fig_result2bc}(b). The analysis reveals that the FS-CF values can be approximated by a normal distribution. As shown in Fig. \ref{fig_result2bc}(b), this distribution has a mean of 0.066 dB and a variance of 0.253 dB$^2$, providing a preliminary quantification for FS-CF values under NLoS conditions.

\subsection{The Accuracy of the Proposed Channel Model}

To further validate the accuracy of the proposed ISAC channel model, which combine the coupling target and background channels, we apply a Similarity Index (SI). The SI has been widely used in MIMO over-the-air studies and standardized performance testing \cite{yuan2022spatial}. Here, it is utilized to quantify the similarity between the measured ISAC channel and the model generated results. The SI calculation in the angular domain is expressed as
\begin{equation}
\begin{aligned}
\label{eqn_si}
S I^{\phi}= & 1-\frac{1}{2} \int\left|\frac{{P}^{\mathrm{mea}}(\phi)}{\int {P}^{\mathrm{mea}}(\phi) d \phi}-\frac{{P}^{\mathrm{model}}(\phi)}{\int {P}^{\mathrm{model}}(\phi) d \phi}\right| d \phi \\
= & 1-\frac{1}{2} \sum\left|\frac{{P}^{\mathrm{mea}}(\phi)}{\sum {P}^{\mathrm{mea}}(\phi)}-\frac{{P}^{\mathrm{model}}(\phi)}{\sum {P}^{\mathrm{model}}(\phi)}\right|,
\end{aligned}
\end{equation}
where ${P}^{\mathrm{mea}}(\phi)$ and ${P}^{\mathrm{model}}(\phi)$ are the power of the measured and generated channels at a specific angle $\phi$. Since the measured CIR has resolution in the delay and angular domains, rather than being continuous, the integration required for SI computation is effectively replaced by a summation. By substituting the variable $\phi$ in (\ref{eqn_si}) with the delay $\tau$, the $SI^{\tau}$ reflects the similarity of the two datasets in the delay domain. Similarly, the $SI^{\tau,\phi}$ can represent the similarity of the datasets in the two-dimensional space of delay and angle. The range of SI is $[0, 1]\times100\%$, where $100\%$ indicates complete similarity, and 0\% represents the total independence.

\begin{figure}[h]
\centering
\subfloat[]{\includegraphics[width=3.3in]{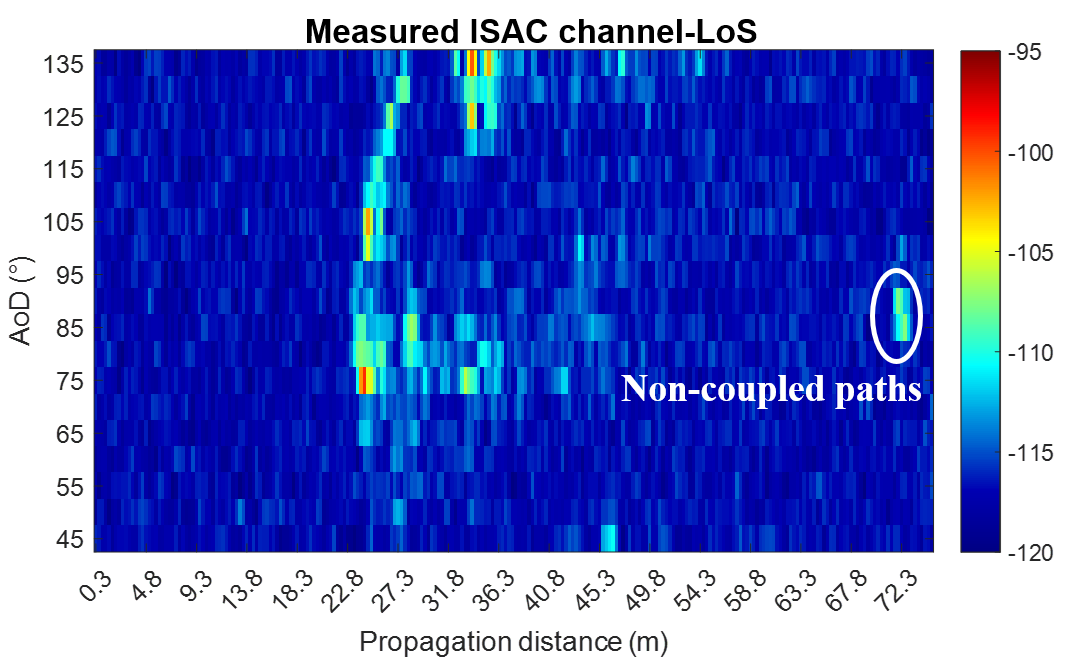}}\\
\vspace{-4mm}
\subfloat[]{\includegraphics[width=3.3in]{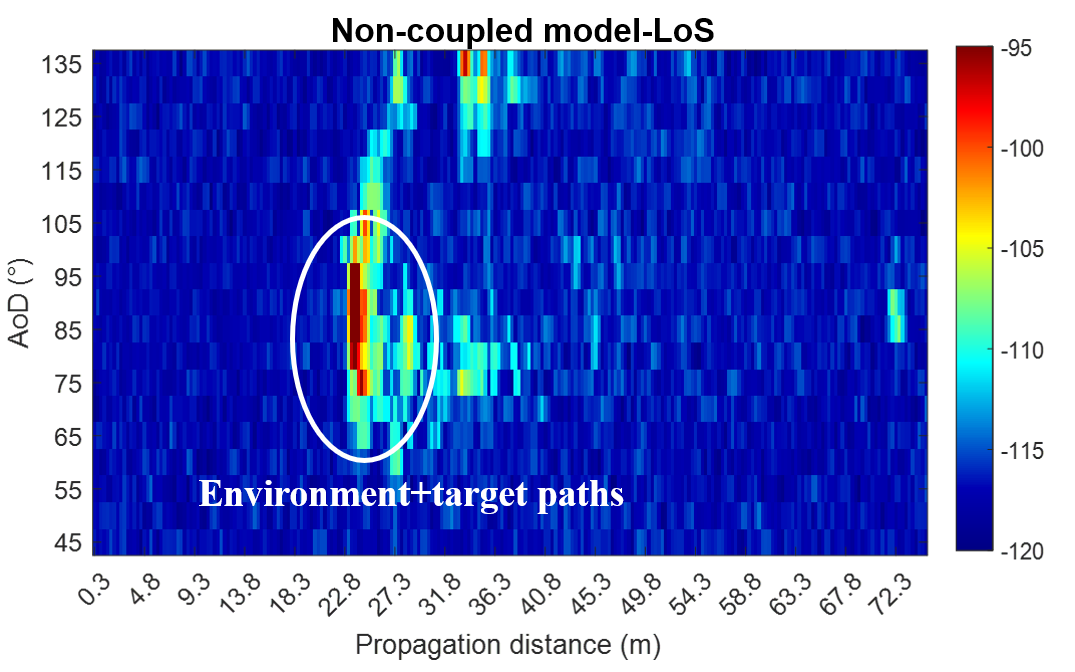}}\\
\vspace{-4mm}
\subfloat[]{\includegraphics[width=3.3in]{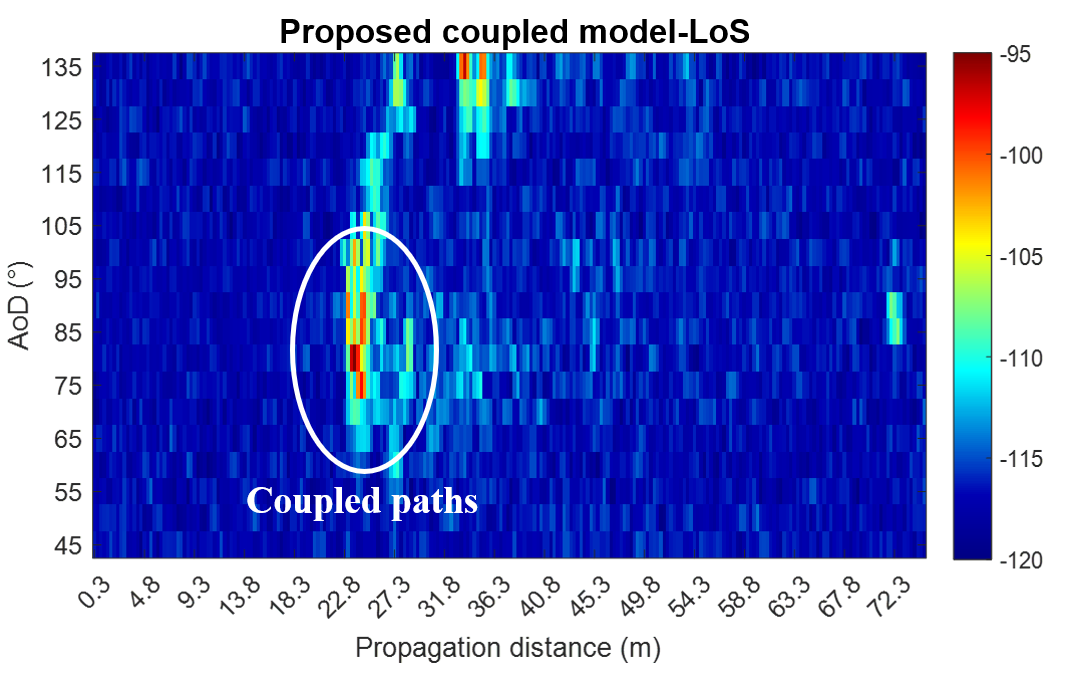}}
\caption{The validation results when the ST is positioned in Point \#2, where (a) measured ISAC channel, (b) ISAC channel by non-coupled model, and (c) ISAC channel by proposed coupled model, respectively.}
\label{fig_validation}
\end{figure}

For demonstration, we analyze validation results at Points \#2 and \#6. Fig. \ref{fig_validation} illustrates the validation results at Point \#2, where the horizontal axis represents the propagation distance, the vertical axis indicates the AoD, and the color scale reflects the received power in dB. Fig. \ref{fig_validation}(a) displays the measured ISAC channel $h_{\text{sen}}$, which corresponds to a magnified portion of Fig. \ref{fig_padp}(b). 
In Fig. \ref{fig_validation}(b), the measured environment channel ${{h}}_{q,p,{e_0}}^{\rm{env}}(t,\tau )$ (from Fig. \ref{fig_padp}(a)) is directly superimposed with the extracted target channel multipaths based on measurements.
A circled part highlights significant power differences compared to the measured ISAC channel in Fig. \ref{fig_validation}(a). This discrepancy arises primarily due to the omission of coupling effect contributed by the interaction between the ST and the environmental LoS paths.

\begin{figure}[h]
\centering
\subfloat[]{\includegraphics[width=3.3in]{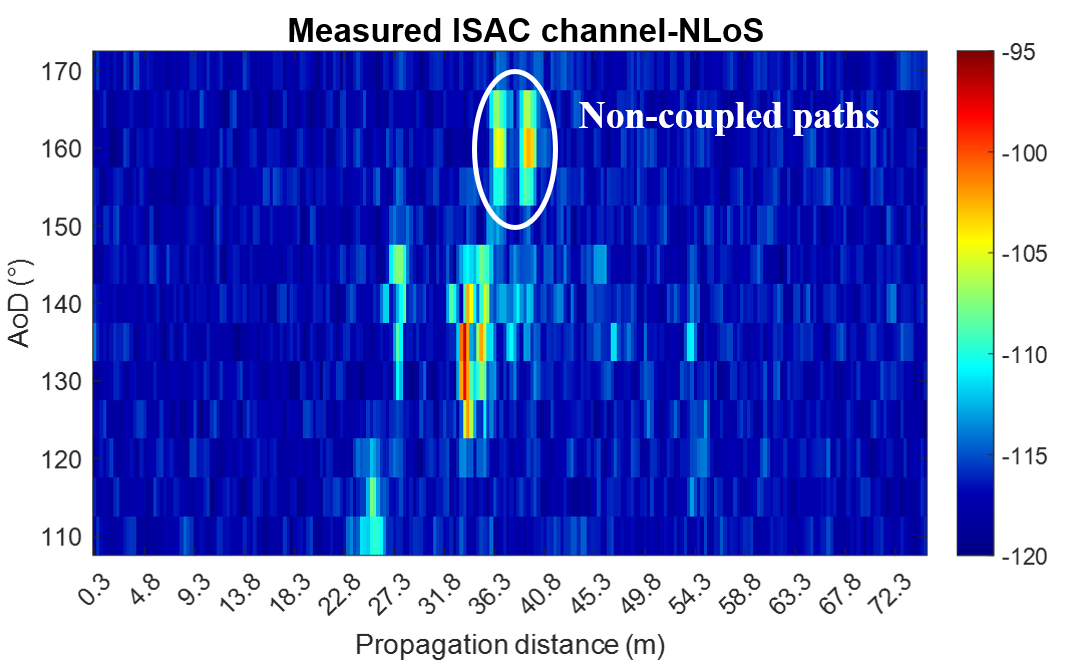}}\\
\vspace{-4mm}
\subfloat[]{\includegraphics[width=3.3in]{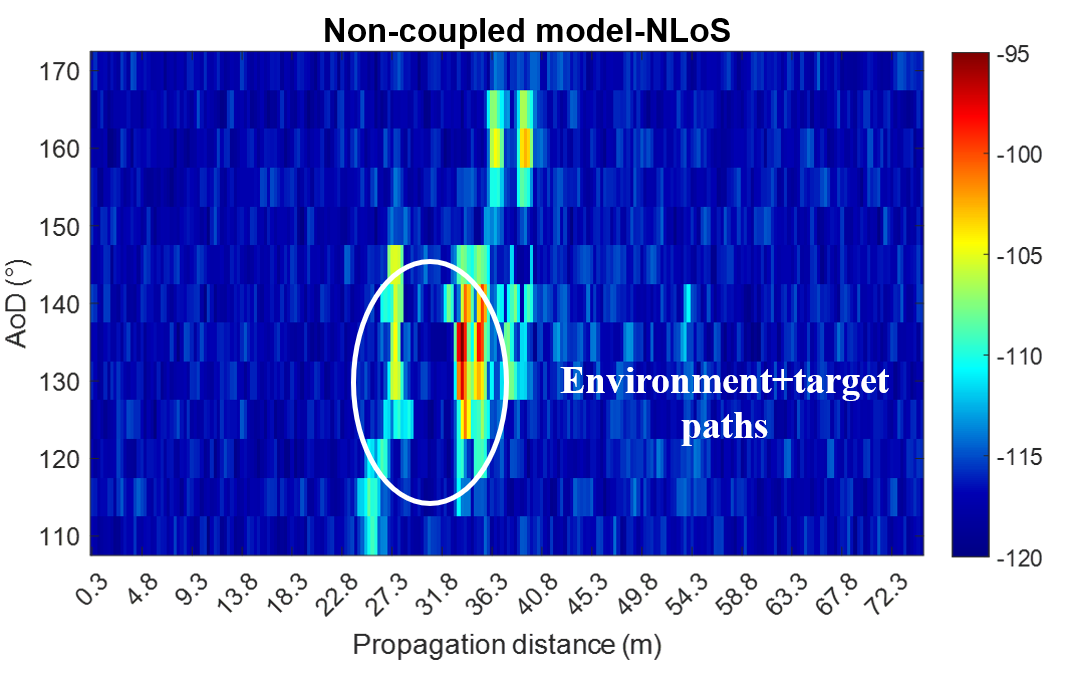}}\\
\vspace{-4mm}
\subfloat[]{\includegraphics[width=3.3in]{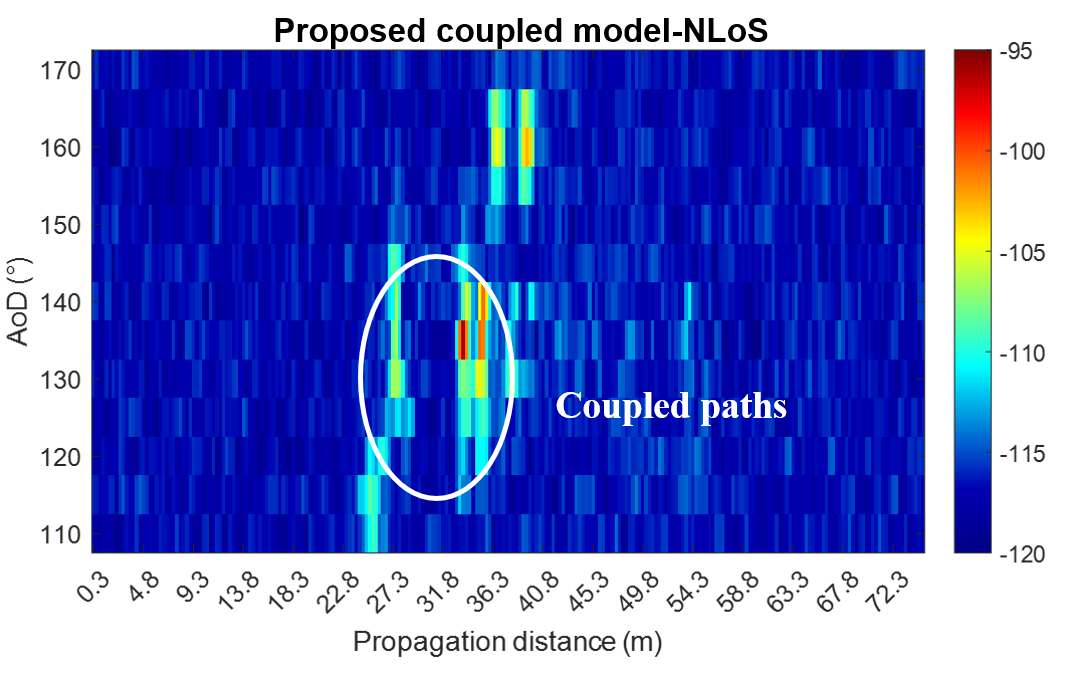}}
\caption{The validation results when the ST is positioned in Point \#6, where (a) measured ISAC channel, (b) ISAC channel by non-coupled model, and (c) ISAC channel by proposed coupled model, respectively.}
\label{fig_validation2}
\end{figure}

Next, we generate the ISAC channel based on the proposed modeling framework in Section \ref{section3} (illustrated in Fig. \ref{fig_frame}). The model parameters are derived from the results in Section \ref{section4}-A), where, within the 70°-110° angular BR and the predefined delay range, the BR-CF values for effective multipaths are set to 0, while the FS-CF values are obtained based on the 4KED-G model. As shown in Fig. \ref{fig_validation}(c), it can be observed that the coupled paths generated by the model effectively reduce the power, making it closer to the measured data in Fig. \ref{fig_validation}(a). To quantify the model accuracy between the conventional non-coupled model (Fig. \ref{fig_validation}(b)) and the proposed coupled model (Fig. \ref{fig_validation}(c)), we calculate their SI values against Fig. \ref{fig_validation}(a), as shown in the first two rows of Table. \ref{table_SI}. The results indicate that the proposed ISAC channel model enhances similarity to the measured ISAC channel by approximately 30\% across all dimensions compared to the non-coupled model. (It should be noted that the delay dimension contains more data, introducing greater uncertainty, which results in an overall SI value lower than that of the angular dimension.) 

The validation results at Point \#6 are illustrated in Fig. \ref{fig_validation2}.
Due to the neglect of the coupling effect arising from the interaction between the ST and the environmental NLoS paths, the non-coupled model exhibits excessive power within the circled part, as shown in Fig. \ref{fig_validation2}(b), compared to the measured data in Fig. \ref{fig_validation2}(a).
The simulated ISAC channel generated using the proposed coupled model incorporates BR-CF values of the effective multipaths set to 0 within 110°-150° and the preset delay range. The FS-CF values follow the normal distribution extracted in Section \ref{section4}-A. As illustrated in Fig. \ref{fig_validation2}(c), the coupled paths generated by the proposed model effectively adjust the power within the circled part. The calculated SI values demonstrate that the proposed model achieves a statistical average gain of approximately 5\%, as presented in the last two rows of Table \ref{table_SI}.

\begin{table}
\caption{SI between measured and modeled ISAC channels\label{table_SI}}
\centering
\begin{tabular}{lccc}
\hline
Values & $S I^{\tau,\phi}$ & $S I^{\tau}$ & $S I^{\phi}$\\
\hline
Non-coupled model - LoS & 48.99\% & 51.72\% & 61.57\%\\
Proposed coupled model - LoS & 79.02\% &  86.55\% & 92.06\%\\
Non-coupled model - NLoS & 84.44\% & 89.02\% & 92.10\%\\
Proposed coupled model - NLoS & 87.16\% &  93.94\% & 97.36\%\\
\hline
\end{tabular}    
\end{table}

\section{Conclusion}\label{section5}
In this paper, we propose a realistic ISAC channel model that effectively captures the coupling effect of STs on the environment. Firstly, we conduct the channel measurement campaign in a typical InF scenario at 105 GHz, using a loaded AGV as the ST. The PADPs demonstrate that the ST couples complexly with environmental scatterers, potentially blocking some multipaths and generating new ones, resulting in power variations compared to the original channel.
To characterize the coupling region, this paper defines BR, centered on the ST and extending beyond its angular size, which is extracted as 40° for both LoS and NLoS measurement scenarios. Based on these observations, we propose a coupled ISAC channel model incorporating two novel parameters: BR-CF and FS-CF. The BR-CF describes path attenuation within the BR in the background channel, while the FS-CF quantifies the power of new, coupled paths introduced by STs in the target channel.
Under LoS conditions, BR-CF and FS-CF values align with Fresnel diffraction theory, while those under NLoS conditions follow a normal distribution.
To validate the proposed model, we introduce an SI to compare the channel generated by conventional non-coupled model and the proposed coupled model against the measured data. The results demonstrate that the coupled model achieves approximately 30\% and 5\% higher accuracy under LoS and NLoS conditions, respectively. This proposed ISAC channel model provides a realistic representation of the ST-environment coupling effect, supporting algorithm design, performance evaluation, and contributing to standardization efforts for ISAC technologies.

\bibliography{reference}

\vfill

\end{document}